\documentclass[journal,onecolumn]{IEEEtran} 
\usepackage{amsmath}
\usepackage{pgfplots}
\usepackage{bbold}
\usepackage[english]{babel}
\usepackage{subcaption}
\usepackage{url}%
\usepackage{cite}%
\usepackage{mathtools}
\usepackage{amssymb}
\usepackage{amsthm}

\newtheorem{theorem}{Theorem}[section]
\newtheorem{lemma}[theorem]{Lemma}
\newtheorem{corollary}[theorem]{Corollary}

\theoremstyle{definition}
\newtheorem{definition}{Definition}[section]

\newenvironment{Pf}{\textbf{Proof:}}
{\hfill$\blacksquare$}

\newcommand{\Set}{\mathcal}

\IEEEoverridecommandlockouts 

\begin{document}
\title{\LARGE \bf {Distortion based Light-weight Security for Cyber-Physical Systems}}
\author{
{Gaurav Kumar Agarwal$^\star$, Mohammed Karmoose$^\star$, Suhas Diggavi, Christina Fragouli, Paulo Tabuada} \\
	Department of Electrical and Computer Engineering, UCLA, Los Angeles, USA\\
	Email: \{gauravagarwal, mkarmoose, suhasdiggavi, christina.fragouli, tabuada\}@ucla.edu
	\thanks{$^\star$Co-first authors. \newline	
		This work was partially supported by NSF grant 1740047, by NSF grant 1705135, by the Army Research Laboratory under Cooperative Agreement W911NF-17-2-0196, and by the UC-NL grant LFR-18-548554.
		A preliminary version~\cite{cdc} of the results in the submitted manuscript was presented at the 57th Conference on Decision and Control (CDC), 2018. This manuscript contains new results not present in the CDC paper (Section-IIIB, Section IV, and Section VI), extended proofs, and additional examples.
	}
}
\maketitle
\IEEEoverridecommandlockouts
\thispagestyle{empty}
\pagestyle{empty}
\begin{abstract}
In Cyber-Physical Systems (CPS), inference based on communicated data is of critical significance as it can be used to manipulate or damage the control operations by adversaries. This calls for efficient mechanisms for secure transmission of data since control systems are becoming increasingly distributed over larger geographical areas. Distortion based security, recently proposed as one candidate for secure transmissions in CPS, is not only more appropriate for these applications but also quite frugal in terms of prior requirements on shared keys. In this paper, we propose distortion-based metrics to protect CPS communication and show that it is possible to confuse adversaries with just a few bits of pre-shared keys. In particular, we will show that a linear dynamical system can communicate its state in a manner that prevents an eavesdropper from accurately learning the state.
%
%
\end{abstract}

\section{Introduction}
Wireless networked environments are a natural host for  a number of cyber-physical control applications, ranging from autonomous cars and drones, to the Internet-of-Things (IoT), to immersive environments such as augmented reality. It is well recognized that wireless networking is essential to realize the potential of new CPS applications, and is equally well recognized that  private and secure exchange of information are necessary and not simply desirable conditions for the CPS ecosystem to thrive. For instance, personal health data in assisted environments, car positions and trajectories, proprietary interests, all need to be protected. This paper introduces a new approach to secure communication in CPS, that aims to distort an adversary's view of a control system's states. In particular, we will show that a linear dynamical system can securely communicate its state {to a trusted party in a manner that prevents a malicious adversary eavesdropping the communication from accurately learning the state.} 
%

Our starting observation is that information security measures (cryptographic and information theoretic secrecy), are not {well-matched} to  CPS applications as they impose unnecessary requirements, such as protecting all the raw data, and thus can cause high operational costs\footnote{Our work focuses on security against passive adversaries -- alternatively referred to in the literature as \textit{CPS Privacy}.}.
{To illustrate this, we start by comparing existing techniques for ensuring CPS privacy, namely \textit{cryptographic} and \textit{information-theoretic} techniques. 
Cryptographic methods rely on computational complexity as a guarantee for the security of the underlying CPS, \textit{i.e.}, the system is secure against computationally-limited adversaries. Cryptographic methods are universal and therefore are easy to integrate in any system under consideration. However, some of their shortcomings are: 1) they do not provide guarantees against adversaries with unlimited computational power (\textit{e.g.}, quantum adversaries), 2) they utilize computationally-heavy encryption/decryption algorithms, and 3) they come at the cost of high overhead on short packet transmissions, therefore increasing delays\cite{wan2016exploiting,Zan2013KeyAA,Trappe,6521318,Alexandru2020}. 
In fact, those techniques have been previously studied in the context of secure CPSs. For example, homomorphic encryption~\cite{7403296,FAROKHI2016163,7799042} and public-encryption systems~\cite{7244402} have been used to provide security of networked control systems.
Alternatively, information-theoretic methods rely on keys: they have low complexity and do not add packet overhead, but require the communicating nodes to share large keys - every communication link needs to use a shared secret key (for a one-time pad) of length equal to the entropy (effectively length) of the transmitted data\cite{shannon1949communication, TGP2018}. These costs accumulate rapidly given that large CPS applications can have dense communication patterns.

Instead, we propose a lightweight approach which provides security guarantees against computationally-capable (i.e., with unlimited computational power) adversaries, and uses small amounts of keys and low complexity operations. The main observation behind our approach is the following. Consider a general encryption scheme which uses a $K$-bit key to encrypt the states of a dynamical system. From an abstract point of view, such a scheme hides the true value of the state among a set of $2^K$ states; without knowing the value of the key, an outside observer of the encrypted state cannot resolve the ambiguity among these fake states -- we refer to this set as the {\it ambiguity set}. General encryption ({\it e.g.}, cryptographic or information-theoretic) schemes aim at increasing the size of the ambiguity set. Differently, in CPS applications, increasing the size of the ambiguity set may not be effective if all of these states are close to each other in a metric space. 
To make this idea concrete, assume an adversary is trying to locate a drone in order to shoot it down with one missile in its possession. If a $K$-bit encryption scheme is used, the adversary would ideally have a set of $2^K$ possible locations for the drone -- any set of $2^K$ possible locations are equivalent from an information-theoretic security point of view. However, if all locations are in close proximity, an adversary can possibly shoot the missile and hit the target regardless of the actual location. On the other hand, a different encryption scheme which uses only $1$ bit of information, but instead carefully chooses the two possible locations to be far apart, would be more secure against such an adversary. We therefore propose a distortion measure in order to capture this idea.}

\begin{figure}[t]
	\centering
		\includegraphics[width=2.0in]{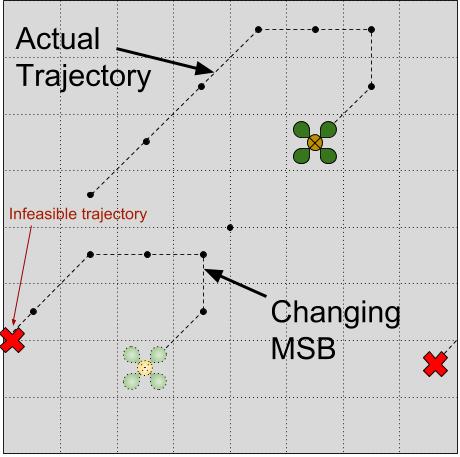}
		\caption{Example of drone motion: protection of the most significant bit.}
		\label{fig::msb}
\end{figure}

The following example illustrates the effect of maximizing distortion\footnote{Although we illustrate our approach for a specific simple example, it extends to protecting general system states.}. Consider  the following simple example of a drone's flying motion, depicted in Fig.~\ref{fig::msb}. The drone starts at any position, and  moves between adjacent points within the grid. It regularly  communicates its location  to a legitimate receiver, Bob. A passive eavesdropper, Eve, wishes to infer the drone's locations, and can perfectly overhear all the transmissions the drone makes. We assume the drone and Bob share just one bit of key, that is secret from Eve, and ask: what is the best use we can make of the key?

Using the one bit of shared key to protect the most significant bit (MSB) is not a good solution. {The} MSB can be protected by
{XORing} a one bit of {shared key with} the MSB.
As shown in Fig.~\ref{fig::msb} the adversary can discover the fake trajectory after a few time steps since this scheme leads to trajectories that do not adhere to the dynamics or environment constraints.
{In particular, the fake trajectory abruptly moves from the left end of the grid to the right end.} At this point, the adversary can learn the real trajectory by flipping back the MSB (we assume that the used scheme is known to everyone). 
Similar attacks can be made if we use a one-time pad~\cite{shannon1949communication} using the same keys over time: as time progresses, more fake trajectories can be discovered and discarded.
%

Conventional entropy measures also fail to provide insights on how to use the  key. For instance, assume we label the $64$ squares in  Fig.~\ref{fig::msb} sequentially row per row, and consider two cases: in case I, Eve learns that  the drone is in one of the neighboring squares $\{1, 2\}$, each with probability 1/2. For case II, Eve knows that the drone is in one of the squares $\{1, 64\}$, again each with probability 1/2. Both cases are equivalent from an information security perspective since in both cases Eve's uncertainty is a set of two equiprobable elements
and hence its entropy is 1. However, the security risks in both situations are different. For example, if Eve aims to take a photo of the drone, in the first case she knows where to turn her camera (squares $1$ and $2$ are close by) while in the second case, she does not (squares $1$ and $64$ are far apart). 

Instead, we propose to use an Euclidean distance distortion measure: how far (in Euclidean space) is Eve's estimate from the actual location. We then propose encoding/decoding schemes which {utilize} the shared key to maximize this distance. We first consider an ``average'' distortion measure. Note that if Eve had not received any of the drone transmissions, then the best (adversarial) estimate of {the} drone's location at any given time is the center point of the confined region in Fig.~\ref{fig::msb}. Therefore, a good encryption scheme would strive to maintain Eve's estimate to be as close to the center point as possible; and we achieve the maximum possible distortion, if, after overhearing the drone's transmissions, Eve's best estimate {still remains} the center point. 

\begin{figure}[t]
		\centering
		\includegraphics[width=2.0in]{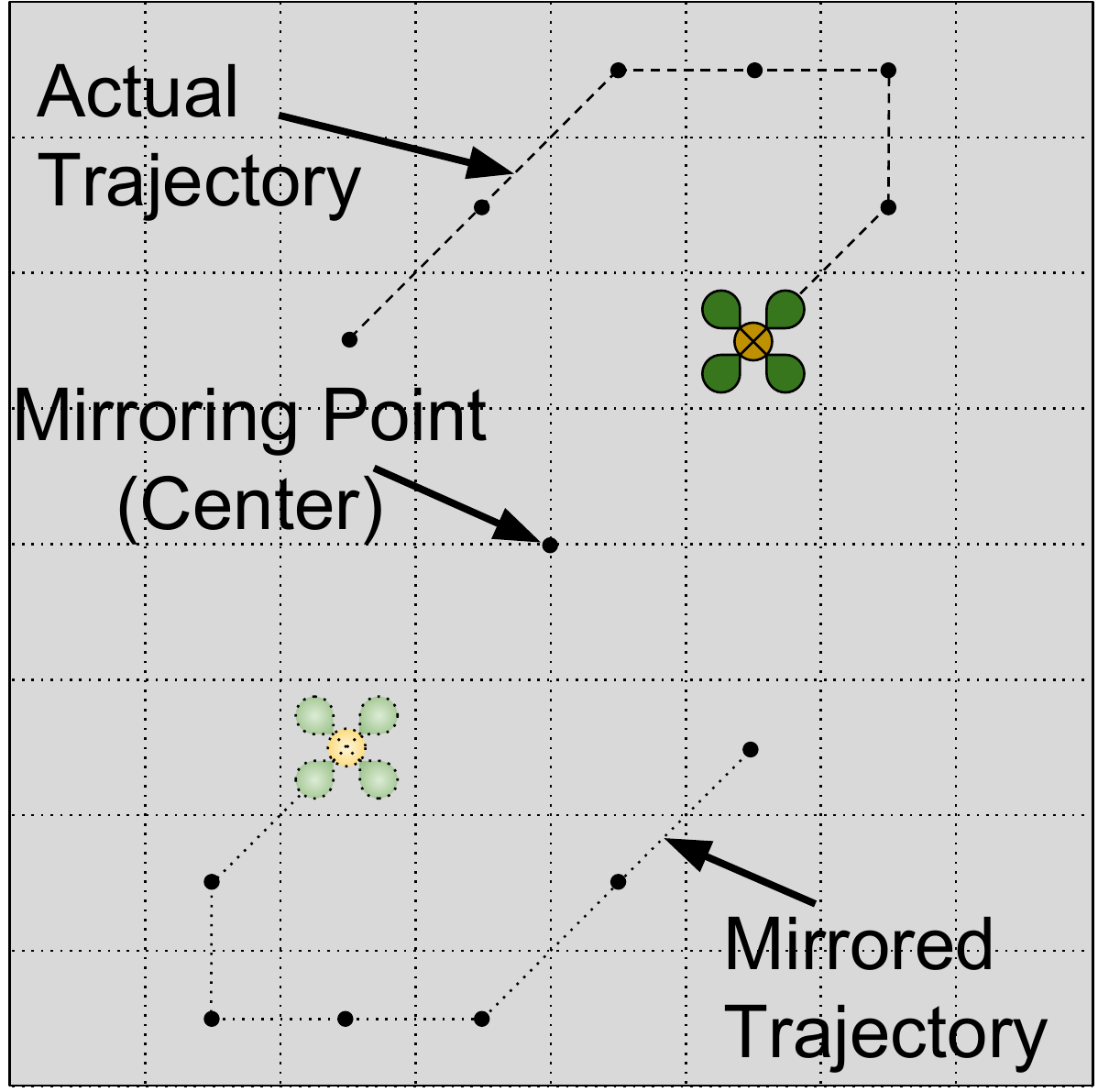}
		\caption{Example of drone motion: mirroring based scheme.}
		\label{fig::mirroring}
	\end{figure}

The following scheme can achieve this maximum distortion by using exactly one bit of shared secret key. When encoding, the drone either sends its actual trajectory, or a ``mirrored'' version of it, depending on the value of the secret key. The mirrored trajectory is obtained by reflecting the actual trajectory across a mirroring point in space; in this example, the mirroring point is the center point {in} Fig.~\ref{fig::mirroring}. Since Eve does not know the value of the shared key, its best estimate of the drone's location - after receiving the drone's transmissions - would be the average location given the trajectory and its mirrored version, which is exactly the center point. 

Our  results in Section~\ref{sec:expected} extend this idea of mirroring to more general light-weight mappings for dynamical systems in higher dimensional spaces, and theoretically analyze the performance in terms of average distortion for a larger variety of distributions (with certain symmetry conditions). We also discuss a {class of systems and controllers} for which we can always achieve the perfect distortion with just one bit of key.

%

Next, we consider a worst-case-sense distortion-based metric. In other words, our security metric is ``in the worst case, how far is Eve's estimate from the actual location?'' That is, the adversary's distortion may be different for different time instances and different instances of the actual trajectory, and we are interested in the minimum among these. For example when the drone is near the origin, both the actual and fake trajectory are close to origin and thus the adversary knows that the drone is near the origin. In this case, the overall (expected) distortion for adversary is still maximum, but at this particular instance of time, the adversary comes very close to the actual position. In Section~\ref{sec:worst} we provide encryption schemes that are suitable for maximizing this distortion metric and show that with $3$ bits of shared key per dimension (\textit{i.e.}, $9$ for three dimensional motion), our schemes achieve near-perfect worst case distortion.
Our main contributions are as follows:

{
\noindent$\bullet$ We} define security measures that are based on assessing the distortion: in the average sense over time and over data, and in the {minimum} sense, providing worst case guarantees at any time and for any particular instances of data. \\
\noindent$\bullet$  For the expected distortion, we develop a scheme which uses exactly one bit of key and can provide maximum possible distortion {(equivalent to Eve with no observations)} in some cases. We also discuss the cases where it is not optimal and give an analytical characterization of the attained distortion. {We then discuss a class of systems and controllers for which we can always guarantee the perfect distortion with just one bit of shared key.} Since for some applications an ambiguity set of size two (corresponding to one bit of key) may not be enough, we also derive an expression of attained distortion when we use larger keys. \\
\noindent$\bullet$ For the worst case distortion, we design a scheme that uses $3$ bits {of key} per dimension and prove it achieves the {maximum possible distortion (equivalent to Eve with no observations)} when the inputs to the systems are independent from the previous states.\\
\noindent$\bullet$ {For linear control systems, we provide a relation between the distortion in inputs to the distortion in states. This is particularly useful when inputs are easier to distort and analyze compared to the states.}

\subsection{Related Work}

Secure data communication where the adversary has unlimited computational power is studied from the lens of information theory, most notably by Shannon~\cite{shannon1949communication} and Wyner~\cite{wyner1975wire}. The study of secure communication while using distortion as a measure of security is relatively new and is first studied by Yamamoto~\cite{yamamoto1988rate}, where the goal is to maximize the distortion of an eavesdropper's estimate on a message, viewed from an asymptotic (in block length) information-theoretic approach. Schieler and Cuff~\cite{schieler2014rate} later showed that, in the limit of an infinite block length $n$ code, only $\log(n)$ bits of secret keys are needed to achieve the maximum possible distortion. The idea of using finite block length (and even single-shot) distortion as a performance measure was initiated in~\cite{chiyo17}, where schemes for single shot communication were considered. It demonstrated the exponential benefits for each additional bit of shared key. The schemes examined were for single-shot sensor observations, and not for time-series data, which is the focus of our work in this paper.

Secure communication in control systems is studied in~\cite{Tsiamis2017StateSecrecyCF, TSIAMIS20178385,tanakaDirected,Malik2013,Pappas16}. Securing the system state from an adversary was explored in~\cite{Tsiamis2017StateSecrecyCF, TSIAMIS20178385}, where an asymptotic steady-state analysis was explored. In contrast our work also deals with transients and is not asymptotic. Information-theoretic security was explored in~\cite{tanakaDirected}, where the mutual information was used as a privacy measure. Security of the terminal state is considered in~\cite{Malik2013} where an adversary makes partial noisy measurement of the state trajectory.
{Securing the states of an unstable system has been considered in~\cite{wiese2018secure} where the notion of secure capacity was used to characterize the level of secrecy against an adversary connected through a wiretap channel.}
Differential privacy for control systems was explored in~\cite{Pappas16}, which uses standard statistical indistinguishability which is equally applicable to categorical (non-metric space) data; in our work, we use the estimation error of the adversary in order to quantify privacy, utilizing the fact that CPS data lies in an  Euclidean space, as argued earlier.
%

\subsection{Notation} For a matrix $A$, we denote by $A^\prime$ {and $A^H$} the transpose {and complex-transpose} of $A$, {respectively};
by $A^{r}$ the $r$-th power of $A$;
{$X$ and $X_a$ denote {column vectors}, and 
$X_{a}^{b} = [X_a^\prime \: X_{a+1}^\prime \: \cdots \: X_b^\prime]^\prime$ for $b\geq a$ and $a,b \in \mathbb{Z}$; $f_X(x)$ denotes the probability density function of a random vector $X$};
for any random vector $Y$, we denote the mean and covariance matrices of $Y$ by $\mu_{Y}$ and $R_{Y}$ respectively, thus for example, the mean and the covariance matrix of $X_a^b$ will be denoted by $\mu_{X_a^b}$ and $R_{X_a^b}$ respectively;  by $[m]$ we denote $\{1,2,\ldots,m\}$ where $m \ \in \mathbb{Z^+}$; and by $[m_1:m_2]$ we denote $\{m_1,m_1+1,\ldots,m_2\}$ where $m_1, m_2 \ \in \mathbb{Z^+}$ and $m_2 > m_1$;
{a negative sign $(-)$ in the superscript of a function indicates the inverse of the function, \textit{i.e.}, the inverse of the functions $\alpha(x)$ and $\alpha^{(K)}(x)$ are $\alpha^-(x)$ and $\alpha^{-(K)}(x)$ respectively.}

\section{System Model}
\label{sec:sysmodel}
\subsection{System Dynamics} 
 
We consider the linear dynamical system,
\begin{align}
{\widetilde{X}}_{t+1} &= A {\widetilde{X}}_{t} + B U_t + w_t,   
& Y_{t} = C {\widetilde{X}}_{t} + v_t, \label{eq:sys_model}
\end{align}
{where ${\widetilde{X}}_t \in \mathbb{R}^n$ is the state of the system at time {$t \in \mathbb{N}$}, $U_t \in \mathbb{R}^m$ is the input to the system at time $t$, $w_t \in \mathbb{R}^n$ is the process noise, $Y_t$ {are the} system observations, and $v_t \in \mathbb{R}^p$ is the observation noise.}
We denote $\widetilde{X}_1^T$ by ${\widetilde{X}}$, ${U_{1}^{T-1}}$ by $U$ and $w_1^{T-1}$ by $w$. 
Based on the initial state ${\widetilde{X}}_1$ and target state ${\widetilde{X}}_T$, the controller computes a sequence of inputs that moves the state from initial state ${\widetilde{X}}_1$ to the target state ${\widetilde{X}}_T$ in $T$ time instances.
{We assume that the system uses the obsevations $Y_{1}^{T}$ to optimally estimate the states ${\widetilde{X}}$. The optimal estimates of $\widetilde{X}$ made by the system are denoted by $X$ -- in the case of {\it perfect observation}, {\it i.e.}, noiseless and observable systems, then $X = \widetilde{X}$.}
%

 \subsection{Communication and Attacker/Defender Models}
At each time instance the system (Alice) transmits information about its state {estimate} to a legitimate receiver, which is referred to as Bob, via a noiseless link. 
This situation occurs for example when Bob is remotely monitoring the execution of the system as in {Supervisory Control And Data Acquisition (SCADA)} systems or in the remote operation of drones.


{\noindent \textbf{Attacker Model.}}
A malicious receiver, referred to as Eve, is assumed to eavesdrop on the communication between the system and Bob and is able to receive all transmitted signals. The goal of Eve is to make an estimate that is as close to $X$ as possible: since Bob receives $X$ and makes control decisions with this information, Eve is interested in $X$. {We assume that Eve knows: 1) the encoding/decoding functions used by Alice and Bob, 2) the dynamical system and 3) the controller design. This information automatically implies knowing the prior probability distributions on the input and state vectors. With this information set, we assume that Eve uses the most adversarial eavesdropping strategy: one which minimizes our performance metric (see Section~\ref{sec::bob_eve}).} Eve is assumed to be passive: she does not actively communicate but is interested in learning the system's states from $t=1$ to $T$.\\
{\noindent \textbf{Defender Model.} We assume that Alice and Bob have a shared {$k$-bit} key $K$ which they use to encode/decode the transmitted messages. For a given encoding/decoding function, the assumed Eve adopts the most adversarial eavesdropping strategy (from the perspective of our chosen performance metric). Therefore, we assume that Alice/Bob attempt to design their encoding/decoding functions which optimize this worst-case performance. We elaborate more on that in Section~\ref{sec::bob_eve}.}
%

 \subsection{Inputs and States Random Process Model}
We assume that both receivers are only aware of the system model, the matrices $A, B, C$ and the statistics of noises. Therefore, from the perspective of the receivers, the input and output sequences have random distributions which depend on $A, B, C$ and the statistics of the noise.  
In addition to the process noise $w$, the joint distribution $f(X,U,w)$ depends on {\it i)} the initial and target states, {\it ii)} the control law of the system and {{\it iii)} the state estimation process}. So, even in noiseless systems, $X$ and $U$ possess inherent randomness from a receiver's perspective due to its lack of knowledge about the initial and target states. 

\subsection{Encoding Model}
{The system encodes and transmits packets $Z_1^T$ to ensure that Bob is able to accurately receive $X_1^T$, the optimal estimates of the system. To do so,} 
the system transmits a packet $Z_t$ at each time step $t$.
%
%
%
%
{In this work, we use} light-weight memoryless encryption schemes. The $t$-th transmitted packet {is} a function of only the current state estimate and the shared keys, thus,
$Z_t \coloneqq \mathcal{E}_t (X_t, K)$, where  $\mathcal{E}_t$ is the encoding function used at time $t$. We will denote $Z_1^T$ by $Z$.\\
\subsection{Bob/Eve Models of Decoding}
\label{sec::bob_eve}
Bob {noiselessly receives the} transmitted packets from the {system, and decodes them using the shared key. Then, using the decoded information, it generates an estimate of the state of the system at times $t \in [T]$.} 
{We require that Bob's estimate is as accurate as Alice's. If we assume that, at time $t \in [T]$, Bob's decoding function is $\Gamma_t\left( Z_1^t,K \right)$, then the previous condition is satisfied by ensuring that $\Gamma_t\left( Z_1^t,K \right) = X_t$ for all $t \in [T]$.}
%
%
%
%
%
%

Similarly, Eve also {receives} all transmissions from the system. {However,} unlike Bob, she does not have the key $K$. 
%
Therefore, Eve's estimate of $X_t$ is $\hat{X}_{t} \coloneqq \Set{\phi}_t \left(Z_1^T\right),  t \in [T]$,
where $\phi_t$ is the decoding function used by Eve at time $t$.

\subsection{Distortion Metrics} We consider a distortion-based security metric which captures how far an estimate is from the actual value. In particular, {our analysis is based on the} Euclidean distance as our distance metric. {However, our analysis can be extended to any $p$-norm,} since other norms are just a constant factor away, \textit{i.e.}, $\|X\|_p \leq n^{\frac{1}{p} - \frac{1}{q}} \|X\|_q$. We assess the performance of Eve as how far its estimate $\hat{X}$, is from Alice's estimate $X$. Formally, for a given time instance $t$ and a transmitted codeword $Z_1^T$, we define the following quantity,
\begin{align}
D(t,Z_1^T) & \coloneqq \mathbb{E}_{X_t|Z_1^T} \left\Vert X_t - \hat{X}_t \right\Vert^2 \stackrel{(a)}{=} \text{tr}\left( R_{X_t|Z_1^T} \right), \label{DE_single}
\end{align} 
where \eqref{DE_single} captures the distortion incurred by Eve while estimating $X_t$ for transmitted symbols $Z_1^T$.
Equality in (a) follows because the best (minimizing) estimates of Eve at time $t$ are, $
 \hat{X}_t  = \Set{\phi}_t \left(Z_1^T\right) = \mathbb{E} \left[X_t|Z_1^T \right].$ 
%
 
Note that Bob is required to successfully estimate $X_t$ knowing ${Z_1^t}$ and the key. Therefore, for a given realization of the key, the encoding function can only map one $X_t$ and that key realization to each value of $Z_1^T$. Therefore Eve realizes {that only trajectories from a {particular subset} can be the true trajectory} for a given $Z_1^T$: those are the ones which correspond to each key realization. Therefore, the expectation in \eqref{DE_single} is in fact taken over the randomness in the key {taking into account} {posterior} probabilities {given} $Z_1^T$. 
{If Eve does not have observations, the expectation is taken} over $X_t$ with prior distribution and we get $D(t,Z_1^T) = \text{tr}(R_{X_t})$.


As $D(t,Z_1^T)$ is a function of time $t$ and the transmitted sequence $Z_1^T$, we consider two overall distortion metrics: the ``average case" distortion (denoted by $D_E$) where we take expectation over all possible $Z_1^T$ and average out over time; and the ``worst case" distortion (denoted by $D_W$) where we take minimum over all possible $Z_1^T$ and time instances.
\begin{align}
&\begin{array}{ll}
\text{Average} \\ \text{Distortion}
\end{array} - \:\:
D_E \coloneqq \mathbb{E}_{Z_1^T} \left[ \frac{1}{T} \sum\limits_{t=1}^T D(t, Z_1^T) \right]  \label{eq:average_case}\\ 
&\begin{array}{ll}
\text{Worst Case} \\ \text{Distortion}
\end{array} - \:\: 
D_W \coloneqq \min\limits_{Z_1^T} \left[ \min\limits_{t\in [T]} D(t, Z_1^T) \right] \label{eq:worst_case} .
\end{align}
%

{It is worth to note that the definitions of $D_E$ and $D_W$ in~\eqref{eq:average_case} and~\eqref{eq:worst_case} imply that Eve's state estimation must be associated to a time instance. In other words,
making a random/constant estimate of the state hoping that it matches the actual state at some time will lead to high distortion values.}
Further, $D_W$ can be defined even when there is no prior distribution on  $X_1^T$. However, {to provide} a baseline comparison {with} the case when the adversary has no observations, we assume that $X_1^T$ always have a known prior distribution.

%

\subsection{Design Goals} Our goal is to choose the encoding and decoding functions, $\mathcal{E}_t$ and $\gamma_t$, so that Bob can decode loselessly while the distortion is maximized for Eve's estimate. In addition, we seek to achieve this with the minimum amount of shared keys $K$. In absence of any observation by Eve, these distortions will be, 
\begin{align*}
D_E^{\text{max}} &=  \frac{1}{T} \sum\limits_{t=1}^T \text{tr}(R_{X_t}), & D_W^{\text{max}} & = \min\limits_{t \in [T]}\text{tr}(R_{X_t}).
\end{align*}
These will serve as upper bounds as,
\begin{align}
D_E &\!=\!\frac{1}{T} \mathbb{E}_{Z_1^T}\! \sum\limits_{t=1}^{T}\!\text{tr}(R_{X_t|Z_1^T})   \stackrel{(a)}{\leq} \frac{1}{T} \sum\limits_{t=1}^{T} \text{tr}(R_{X_t}) = D_E^{\text{max}},\label{eq::UpperBound} \\
D_W &\!= \min\limits_{Z_1^T}  \min\limits_{t\in [T]} \text{tr}(R_{X_t|Z_1^T}) \leq   \min\limits_{t\in [T]}\mathbb{E}_{Z_1^T} \left[ \text{tr}(R_{X_t|Z_1^T})\right] \nonumber \\
& \stackrel{(b)}{\leq} \min\limits_{t\in [T]} \text{tr}(R_{X_t}) = D_W^{\text{max}} \label{eq::UpperBoundDW} ,  
 \end{align}
{where (a) and (b) follows from $\mathbb{E}_{Z_1^T} \left[  \text{tr}\left( R_{X_t} | Z_1^T  \right) \right]    \leq \text{tr}\left( R_{X_t} \right)$ which follows from the law of total variance.} 
%
%

%

\section{Optimizing Average Distortion $D_{E}$}
\label{sec:expected}

In this section, we will first discuss schemes to optimize the Average Distortion ($D_E$).
{We will initially analyze encoding schemes which use {\it one} bit of secret key, and characterize their attained level of distortion. We then show that such schemes attain the maximum level of distortion for a family of distributions on $X$ which exhibit a certain class of symmetry. Later we describe how this analysis extends to the use of multiple keys, as for some application having an ambiguity set of size two might not be enough.} 

\subsection{Encoding Schemes with $1$-bit Shared Secret Key}
We now discuss {encoding schemes} that use one bit of shared key and show how the achieved distortion compares to the upper bound in~\eqref{eq::UpperBound}. 
{These encoding schemes work as follows:}


\begin{equation}
Z_t = \left\{\begin{matrix}
X_t & \text{if } K = 0, \\
{\alpha_t({X}_t)} & \text{if } K=1,
\end{matrix}
\right. \ \ \forall t \in [T], \label{eq::onebit_scheme}
\end{equation}
where $K \in \{0,1\}$ is the shared bit and {$\alpha_t(X_t)$ is a transformation of} the state vector {$X_t$.} We will next show the attained distortion of such schemes.

\begin{theorem}[Proof in Appendix~\ref{app::MirroringPerformanceCausal}]
\label{thm::MirroringPerformanceCausal}
The average distortion ($D_E$) attained by using the scheme in~\eqref{eq::onebit_scheme} is,
\begin{equation}
\frac{1}{2T} \sum\limits_{t=1}^T \mathbb{E}_{X} \left\{\frac{f_{X}({\alpha^{-}(X))} }{f_{X}({X})+f_{X}({\alpha^{-}(X)})} \left\| {X_t} - {\alpha_t^{-}({X_t}) }\right\|^2 \right\}, \label{eq::onebit_dist}
\end{equation}
where $f\alpha^{-}(X) \coloneqq [\alpha_1^-(X_1)^\prime \: \alpha^-_2(X_2)^\prime \: \cdots \: \alpha^-_T(X_T)^\prime ]^\prime$. Moreover, if the following condition holds,

\begin{align}
f_X(x) &= f_X({\alpha^{-}(x)}), & \text{for all } x \in \mathcal{X}, \label{eq::sym_cond_general}
\end{align}
then the expression simplifies to
\begin{equation}
D_E = \frac{1}{4T} \sum\limits_{t=1}^T \mathbb{E}_X \left\| X_t - \alpha_t(X_t) \right\|^2. \label{eq::onebit_dist_sym}
\end{equation}

\end{theorem}

\begin{figure}
	\centering
	\includegraphics[width=2.0in]{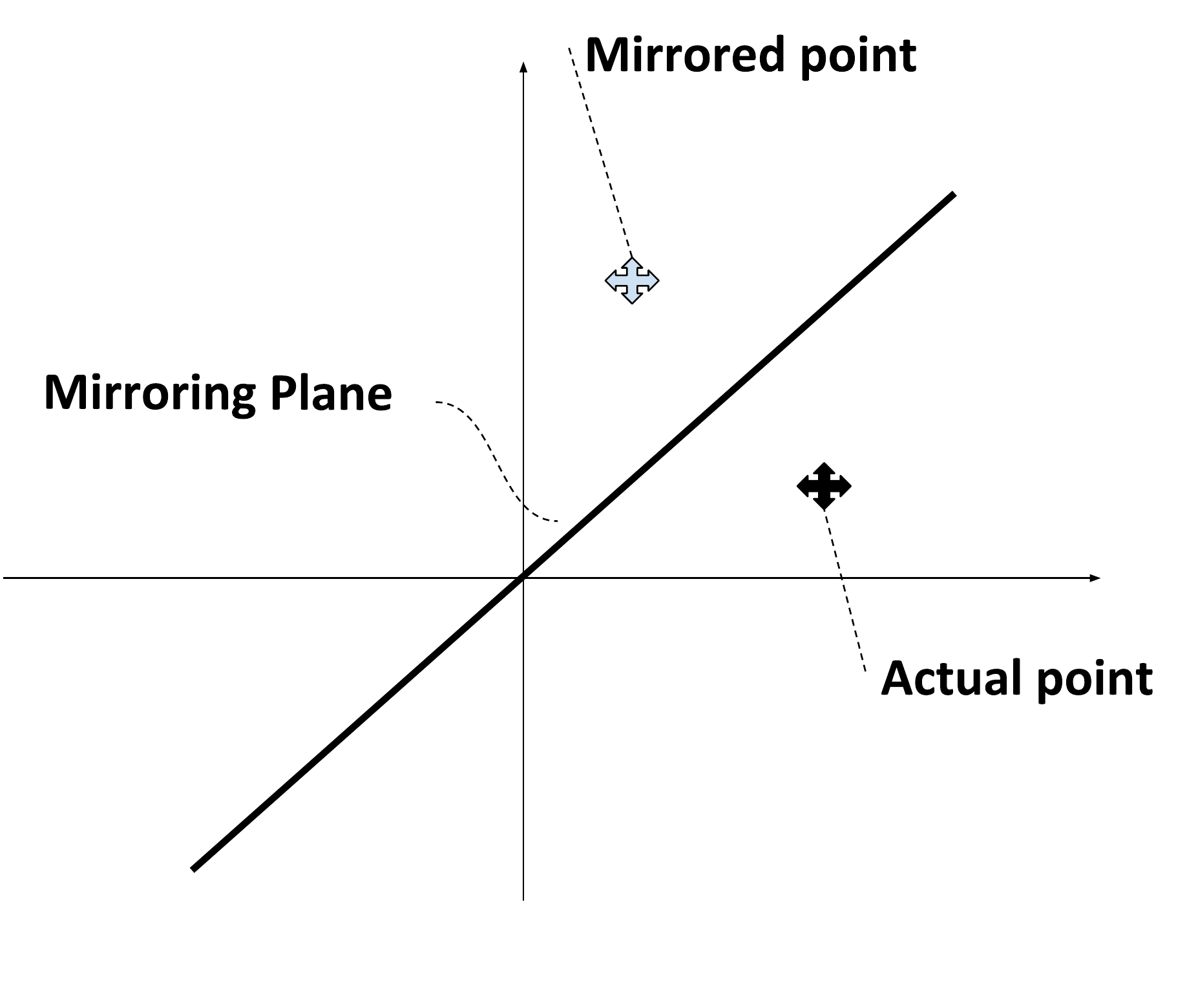}
	\caption{Mirroring across the line passing through the origin and having a $45^\circ$ angle with the $X$-axis.}
	\label{fig:reflection} 
\end{figure}

Condition~\eqref{eq::sym_cond_general} implies a general notion of symmetry in the distribution of $f_X(x)$. In the following, we focus on a particular notion of distribution symmetry,
for which we show the corresponding choice of $\alpha_t(X_t)$ and how it can achieve high levels of distortion. Consider a transformation function $\alpha_t(x)$ which reflects a point $x$ across an affine subspace of dimension $d$, defined by the equations $S_tx = b_t$ where $S_t \in \mathbb{R}^{d \times n}$ consists of $d \leq n$ orthonormal rows, and $b_t \in \mathbb{R}^d$; the transformation is $\alpha_t(x) = \left(I - 2S_t^\prime S_t \right)x + 2S_t^\prime b_t$.
{The choice of the dimension $d$ and the subspace ($S_t, b_t$) depend on the properties we would like the encoded trajectories to have.}
We refer to encoding schemes that are based on this transformation as {\it mirroring schemes}. For example, consider $X_t \in \mathbb{R}^2$ where $S_t = \frac{1}{\sqrt{2}}[-1 \:\: 1]$ and $b_t = 0$. Then {$\alpha_t({X}_t)$} corresponds to mirroring across a line that passes through the origin with a $45^\circ$ angle. This is shown in Fig.~\ref{fig:reflection}. We are interested in {\it mirroring schemes} as they are light-weight and can be implemented on low-complexity {IoT} devices. Moreover, such schemes can provide the maximum distortion level for a class of distributions with what we refer to as {\it Point Symmetry}.


\begin{definition}[\textbf{Point Symmetry}]
A random vector $X$ is said to have Point Symmetry if there exists a point $v$ for which $f_X(x) = f_X(2v-x), \ \forall x \in \mathcal{X}$.
\end{definition}

\begin{lemma}
\label{lem::PointSym}
 If $X$ has Point Symmetry across $v$, then $v = \mu_X$.
\end{lemma}
\begin{Pf}
 Since $X$ has Point Symmetry, then 
 \begin{align*}
 & &f_X(x) &= f_X(2v - x) &\Rightarrow & & f_X(x) &= f_{2v - X}(x) \\
  \Rightarrow &   &\mu_X &= 2v - \mu_X &\Rightarrow &  & \mu_X &= v.
 \end{align*}
\end{Pf}

The following result characterizes the performance of the mirroring scheme, and shows that it achieves the maximum distortion for distributions with Point Symmetry.
%
%

\begin{corollary}
\label{cor::SymmetricDist}
{If $\alpha_t(X_t)$ is based on a \textbf{mirroring scheme} along the planes given by $S_t x = b_t, \ t\in [T]$ and the condition~\eqref{eq::sym_cond_general} holds,
then~\eqref{eq::onebit_dist_sym} becomes,}
%
\begin{equation}
\label{eq::DEE_simple_causal}
D_E = \frac{1}{T} \sum\limits_{i=1}^T \text{tr}\left( S_t R_{X_t} S_t^\prime + (b_t-S_t \mu_{X_t})( b_t - S_t \mu_{X_t})^\prime \right).
\end{equation}
Moreover, if {$X$} has Point Symmetry, then $D_E = \frac{1}{T} \sum\limits_{t=1}^T tr(R_{X_t})$, the maximum possible distortion.
\end{corollary}%
\begin{Pf}
If condition~\eqref{eq::sym_cond_general} holds, then by simply plugging the expression of $\alpha_t(X_t)$ for the mirroring scheme along $S_tx=b_t$ that is $\alpha_t(X_t) = \left(I - 2S_t^\prime S_t \right){X_t} + 2S_t^\prime b_t$ in~\eqref{eq::onebit_dist_sym} we get~\eqref{eq::DEE_simple_causal} (Formal proof in Appendix~\ref{app::MirroringPerformanceCausal}). Choosing $S_t = I $ and $b_t = \mu_{x_t}$ makes ${\alpha^{-}} (X_1^T) = 2 \mu_{X_1^T} - X_1^T$ which by Point Symmetry satisfies~\eqref{eq::sym_cond_general}.
Therefore, we get $D_E = \frac{1}{T} \sum\limits_{t=1}^{T} tr (R_{X_t})$.
Note that the optimal distortion, denoted as $D_E^\star$ and obtained by optimizing over $S_t$ and $b_t$, satisfies $D_E^\star \geq \frac{1}{T} \sum\limits_{t=1}^{T} tr (R_{X_t})$. However, from \eqref{eq::UpperBound}, we have $D_E^\star \leq \frac{1}{T} \sum\limits_{t=1}^{T} tr (R_{X_t})$. Therefore, the selected $S_t$ and $b_t$ and the corresponding distortion value are optimal.
\end{Pf}

Now, we show the implications of our results for mirroring based schemes in the context of a few examples.\\
%
%

\textbf{Example 1.} 
Consider an example where $U$ is distributed as Gaussian with mean $\mu_U$ and covariance matrix $R_U$. Then {for a noiseless system with perfect observation and a zero initial state, $X_2^T$} is also Gaussian distributed with mean ${\mu_{X_2^T}} = Q \mu_U$ and variance ${R_{X_2^T}} = Q R_U Q^T$ {where $Q$ relates $U$ to $X_2^T$ after unfolding the time-dependent state equations into the form $X_2^T = Q U$}. A Gaussian random vector has Point Symmetry and therefore, according to Corollary \ref{cor::SymmetricDist}, we can get maximum distortion by setting $b_t = \mu_{X_t}$ and $S_t = I$.
	
The next example is based on a Markov-based model for the dynamical system. For this example, the following lemma is useful.
	
\begin{lemma}
\label{lem::MarkovModel}
Consider the random vector $X_1^T$ where the following conditions hold: 1) $f_{X_1}(x_1)$ has Point Symmetry, and 2) $f_{X_t|X_1^{t-1}}(x_t | x_1^{t-1})$ has Point Symmetry, then so does $f_{X}(X)$, where $X = X_1^T$ and $\mu = [{\mu_{x_1}}^\prime \:\: {\mu_{x_2}}^\prime \:\: \cdots \:\: {\mu_{x_T}}^\prime]^\prime$. Therefore, by virtue of Corollary \ref{cor::SymmetricDist}, mirroring schemes can achieve the maximum distortion.
\end{lemma}
	
Lemma \ref{lem::MarkovModel} allows us to characterize the performance of the following example.
	
\textbf{Example 2.} Consider {the following random walk} mobility model. Let $a \in \mathbb{N}^+$, and $X_t$ be its location at time $t$, then,
	
\begin{align*}
X_1 &\sim \text{Uni}([-a:a]) \\
X_{t}|X_{t-1} &\sim \text{Uni}([-a:a] \cap \{X_{t-1} -1, X_{t-1}, X_{t-1}+1 \}).
\end{align*}

{This example follows the system model in~\eqref{eq:sys_model} by assuming a noiseless system with $U_t$ to be independent across $t$, and to be uniformly distributed among $\{-1,0,1\}$ when $X_t \in [-a+1:a-1]$, $U_t$ uniformly distributed in $\{0,1\}$ when $X_t = -a$ and $U_t$ uniformly distributed in $\{-1,0\}$ when $X_t = a$.}
One can see that these distributions satisfy the conditions in Lemma \ref{lem::MarkovModel}. Therefore, one can set $b_t = \mu_t = 0$ and $S_t = 1$, which will achieve maximum distortion of $D_E$.

\textbf{Example 3.} Here we provide a numerical example which shows how our mirroring scheme performs for situations where we compute the state distributions using numerical simulations.  {In the Section~\ref{sec:maintain_symmetry}, we will also show that the controller used in this Example falls under the class of controller where we do not need to compute the distribution on states and can directly apply our scheme to achieve the perfect distortion.} We consider the quadrotor dynamical system provided in (4) of~\cite{kumar2012opportunities}. The quadrotor moves in a 3-dimensional cubed space with a width, length and height of 2 meters, where the origin is the center point of the space. The quadrotor starts its trajectory from an initial point $(-1,y_1,z_1)$ and finishes its trajectory at a target point $(1,y_T,z_T)$ after $T$ time steps, where {the points $y_1,z_1,y_T,z_T$ are picked uniformly at random in $[-1,1]^4$}. We assume that $T=10$ time steps, and that the continuous model in~\cite[$(4)$]{kumar2012opportunities} is discretized with a sample time of $T_s = 0.5$ seconds. We assume that the quadrotor encodes and transmits only the states which contain the location information {(first three elements of the state vector $X_t$)}. The quadrotor is equipped with an LQR controller which designs the input sequence $U_1^{T-1}$ as the solution of the following problem
%
 \begin{equation}
 \begin{array}{ll}
  \text{minimize} & {\left\| U \right\|^2 + 10 \left\| X_2^{T-1} \right\|^2} \\
\text{subject to } &  {X_{t+1} = A^{\text{quad}} X_t + B^{\text{quad}} U_t, \quad \forall t \in [T-1] }\\
  & X_1 = \left[ \begin{matrix}
           -1 & y_1 & z_1 & 0 & \cdots & 0 
          \end{matrix}\right]^\prime, \\
  & X_T = \left[ \begin{matrix}
           1 & y_T & z_T & 0 & \cdots & 0 
          \end{matrix}\right]^\prime,
 \end{array}
\end{equation}
where $A^{\text{quad}}$ and $B^{\text{quad}}$ define the quadrotor's discrete-time model. The remaining states of $X_1$ and $X_T$ are set to zero to allow the drone to hover at the respective locations.
We perform numerical simulation of the aforementioned setup: we run {$2$ millions} iterations, where in each iteration a new initial and target points are picked, and the resultant trajectory is recorded. Based on the recorded data, we consider different mirroring schemes and numerically evaluate the attained distortion. To facilitate numerical evaluations, the simulation space is gridded into bins with $0.2$ meters of separation, and the location of the drone at each trajectory is approximated to the nearest space bin. 

\begin{figure}
 \centering
 \includegraphics[width=2.85in]{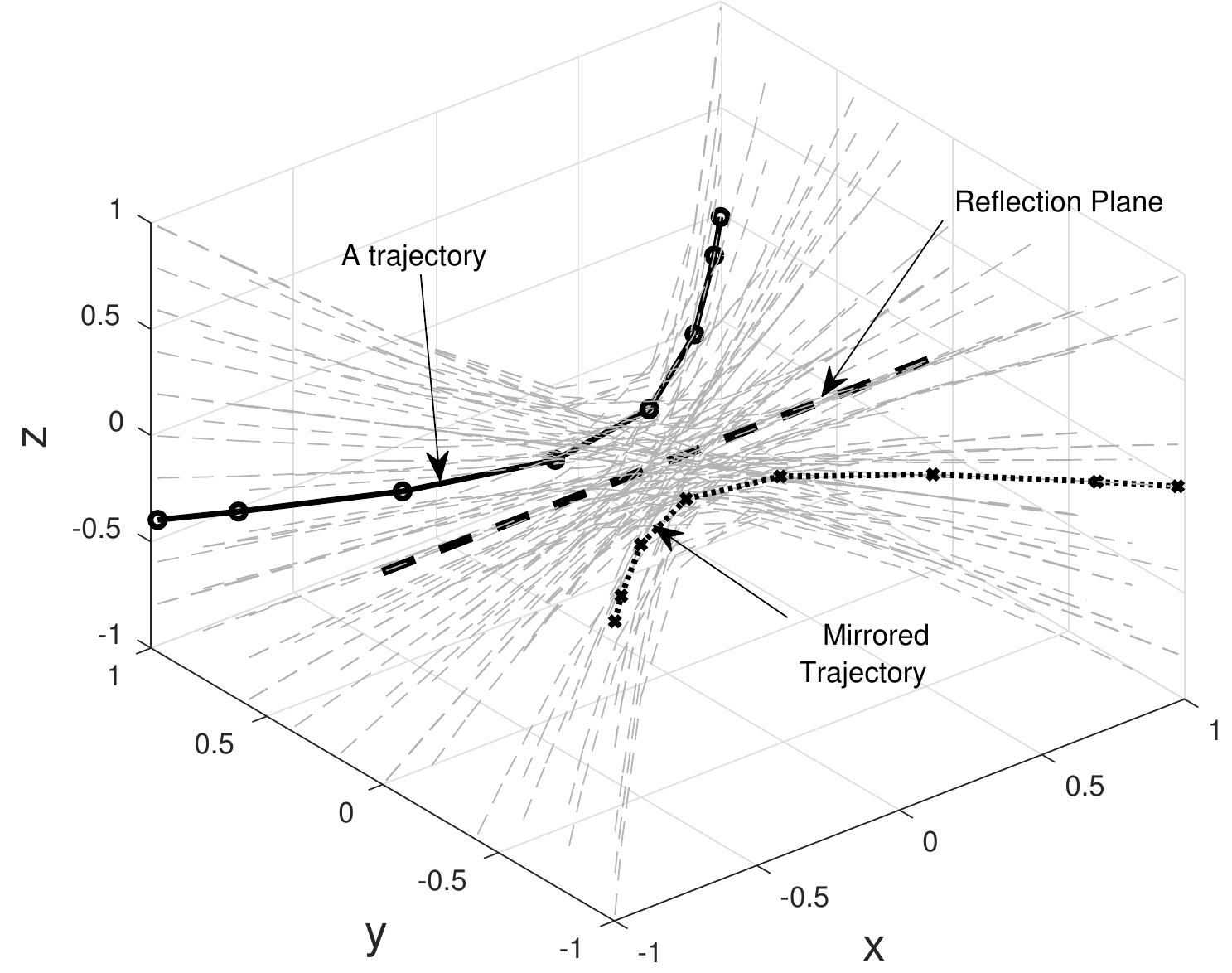}
 \caption{An illustration of some trajectories. The reflection plane is shown as a dashed-black line. One trajectory (solid-black) is shown along with its mirrored image (dotted-black).}
 \label{fig::simulationMirroring}
\end{figure}

Figure \ref{fig::simulationMirroring} shows some of the drone trajectories obtained from our numerical simulation. It is clear that not all trajectories are equiprobable, and therefore the distribution of $X_t$ is not uniform across all bins in space. 
Since the motion of the drone is mainly progressive in the positive x-axis direction, reflection across a fixed point results in mirrored trajectories that are progressing in the opposite direction, and therefore are identified to be fake automatically. Therefore, mirroring across a point here is useless: the numerically computed distortion for this scheme is equal to zero.

Next we consider mirroring across the reflection plane shown in Figure~\ref{fig::simulationMirroring}, where $b_t = 0$ and $S_t = \left[ \begin{matrix}0 & 1 & 0 \\ 0 & 0 & 1 \end{matrix}\right]$. As can be seen from the figure, the reflection plane is indeed an axis of symmetry for the distribution of the drones trajectories, and therefore is expected to provide high distortion values. We numerically evaluate the attained distortion using the scheme by using equation \eqref{eq::onebit_dist}, which evaluates to $D_E = 0.3971$. This is slightly less than $D_E^{\max} = 0.3979$.

\subsection{Encoding Schemes with $k$-bits Shared Secret Key}
The scheme in~\eqref{eq::onebit_scheme} assumes the use of one bit for encryption. However, it is straightforward to extend the scheme when we require a larger ambiguity set. For {$k$} bits, we denote the possible values of the shared key as {$K \in [0:2^k-1]$}. Therefore, the scheme works as follows

\begin{align}
Z_t(K) &= \alpha_t^{(K)}(X_t), \  \forall t \in [T], 
\label{eq::multibit_scheme}
\end{align}
where $\alpha_t^{(K)}$ is an invertible  transformation function used at time $t$ when the value of the key is $K$, and $\alpha_t^{(0)}(x) = {\alpha_t^{-(0)}(x)} = x $. The following theorem shows the achieved value of the distortion in this case, which is a direct extension of Theorem~\ref{thm::MirroringPerformanceCausal}.

\begin{theorem}[Proof in Appendix~\ref{app::multipleKproof}]
 \label{thm::MirroringPerformanceCausal_multibit}
 The average distortion $D_E$ attained by using the scheme in~\eqref{eq::multibit_scheme} is
\begin{align}
 & \frac{1}{2^k T} \sum\limits_{t=1}^T \mathbb{E}_{X} \left\{\frac{\sum\limits_{K=0}^{2^k-1} f_{X}( \alpha^{-(K)}(X)) \left\| R_t^{(K)} \right\|^2}{\left[ \sum\limits_{K=0}^{2^k-1} f_X(\alpha^{-(K)}(X)) \right]^2 f_{X}(X)} \right\},  \label{eq::multibit_dist}
\end{align}
where $R_{t}^{(K)}\!=\!\sum\limits_{\ell=0}^{2^k-1} f_X(\alpha^{-(\ell)}(X)) \left( \alpha_t^{-(\ell)}(X_t) - \alpha_t^{-(K)}(X_t) \right)$ and $\alpha^{-(K)}(X) \coloneqq [\alpha^{-(K)}_1(X_1)^\prime \: \alpha^{-(K)}_2(X_2)^\prime \: \cdots \: \alpha^{-(K)}_T(X_T)^\prime ]^\prime$. Moreover, if the following condition holds,
\begin{align}
 f_X(\alpha^{-(K)}(x)) &= C(x), & \forall x \in \mathcal{X},  \forall K \in [0:2^k-1], \label{eq::sym_cond_general_multibit}
\end{align}
wehre $C(x)$ is a constant, then $D_E$ simplifies to
\begin{align}
&\frac{1}{2^{3k} T} \sum\limits_{t=1}^T \mathbb{E}_{X}\left[ \sum\limits_{K=0}^{2^k-1}\left\|\sum\limits_{\ell =0}^{2^k-1}\left( \alpha_t^{-(K)}(X_t) - \alpha_t^{-(\ell)}(X_t) \right) \right \|^2   \right]. \label{eq::kbit_dist_sym}
\end{align}
\end{theorem}

{
Theorem~\ref{thm::MirroringPerformanceCausal_multibit} shows the average distortion attained for general schemes which use $k$ bits. In addition, we can generalize the mirroring scheme in Section~\ref{sec:expected} to utilize $k$ bits as follows.
Given a $k$-bit key, then we select a set of $k$ hyperplanes (\textit{i.e.}, a set of $k$ parameters, $S_t^{(K)} \in \mathbb{R}^{d \times n}$ and $b_t^{(K)} \in \mathbb{R}^d$) for each time step $t$. Then, let $\mathcal{K}$ be a set of binary values corresponding to the binary representation of the $k$-bit key. The mirroring scheme would transform the point $x$ to $\prod\limits_{K \in \mathcal{K}} (I - 2 {S_t^{(K)}}^\prime S_t^{(K)} )x + 2{S_t^{(K)}}^\prime b_t$, \textit{i.e.}, $x$ is mirrored across the hyperplanes corresponding to the $1$-valued bits in the binary representation of the shared key. It is not difficult to see that this scheme can achieve the maximum distortion when $X_t$ is Gaussian distributed with zero mean and covariance matrix $R_t$ and independent across $t$: for this case, $S_t^{(K)}, K \in \mathcal{K}$ are chosen as the eigenvectors of $R_t$ and $b_t^{(K)} = 0$.
}

{Using multiple bits of shared keys can provide benefits beyond having a larger ambiguity set. In fact, while we show the optimality of 1-bit mirroring schemes for distributions with point symmetries, using multiple bits of shared key can provide a better distortion for general distributions. For example,} it was shown that, for a general finite alphabet: (1) $1$-bit schemes are not sufficient to achieve the maximum distortion, {and} (2) with just 5 bits of shared keys, a scheme achieves more than 97\% of the maximum possible distortion~\cite{chiyo17}. 

\section{Transformations Maintaining Point Symmetry}
\label{sec:maintain_symmetry}
Encoding and decoding schemes such as the ones mentioned in Section~\ref{sec:expected} can be generally used for any dynamical system with arbitrary distributions on the inputs $U$, the state vectors $\widetilde{X}_t$ and the state estimates $X_t$. However, characterizing the attained level of average distortion (using expressions~\eqref{eq::onebit_dist} and~\eqref{eq::multibit_dist}) requires the knowledge of the distribution of the state estimate. While a distribution can be obtained for the initial and target state vectors, it may be difficult to incorporate the system dynamics, the estimation method as well as the controller into the process of finding a distribution of the inputs, {states} and {states} estimate. In such cases, numerical evaluations can aid into finding the needed distribution, as was shown in Example 3 in Section~\ref{sec:expected}. 
Although it is necessary to find the {{state} distribution in order to characterize the distortion, the knowledge of existing {{symmetries} in the distribution can directly give {{possible choices for} the transformation function $\alpha_t(\cdot)$ which {{may attain} high levels of distortion{; for} example, if there is a point symmetry in the distribution, mirroring across the symmetry point attains the maximum possible distortion.
In this section, we ask the following question: ``{\it under which conditions on the dynamical system, does point symmetry in the initial and target states results into point symmetry on the {states} estimate}?''

A general answer to the aforementioned question appears to be difficult. Therefore, we limit our answer in this work to the scope of linear controllers.
{For a given} initial and target state, let $X^{(\text{ref})}$ be the reference trajectory that the control system ideally wishes to follow. We assume that the system controller selects an input vector that is a linear function of $X^{(\text{ref})}$. In many cases, $X^{(\text{ref})}$ is also a linear function of the initial and target states ({\it e.g.}, when the reference trajectory is the solution of an LQR problem for the noiseless version of the system). Then we can write {$ U_t = K_t \left(X_t - X_t^{(\text{ref})}\right)$.}
Moreover, we assume that the optimal estimation function that the system uses is a linear one in the observations, {\it i.e.}, we assume $X_t$ is a linear function of $Y_1^t$, $X_{\text{init}}$ and $X_{\text{target}}$.
By incorporating the controller and estimation equations into the system dynamics, one can arrive at the following relation {$X = MQ$, where $Q =  \left[X^\prime_{\text{init}} \:\:\: X^\prime_{\text{target}} \:\:\: {w_{1}^T}^\prime \:\:\: {v_{1}^T}^\prime \right]$, and}
the matrix $M$ is a function of the matrices $A$, $B$, $C$, $K_t$ and the linear function used in the estimation of state $X_t$ from the observations. We assume that $w_1^T$ and $v_{1}^T$ are uncorrelated Gaussian random vectors. We first prove the following lemma.

\begin{lemma}
\label{lem:symmetry_map}
If {a random vector $V_1 \in \mathbb{R}^n$} has {Point Symmetry across $\mu_{V_1}$}, and $g$ is an affine function, then the {random vector} $V_2 = g(V_1)$ {has} Point Symmetry {across} $g(\mu_{V_1})$.
\end{lemma}

\begin{IEEEproof}
{If $V_1$ has Point Symmetry, then the following conditions are equivalent:}
\begin{align*}
f_{V_1}(v_1) & = f_{V_1}(2\mu_{V_1} - v_1), & \forall v_1 \in V_1, \\
f_{V_1}(v_1) & = f_{2\mu_{V_1} -V_1} (v_1),  & \forall v_1 \in V_1.
\end{align*}

{Thus}, to prove that $V_2$ also has Point Symmetry, it suffices to prove that the density of $V_2$ and $2 \mu_{V_2} - V_2$ is the same. 
{Consider the} two random vectors $W_1$ and $W_2$. {If they have the same support and the same density function}, then $g(W_1)$ and $g(W_2)$ will also have the same density for any {function} $g$; {we denote this by writing $W_1 \sim W_2$}. Thus,
\begin{align*}
V_1 & \sim  2 \mu_{V_1} - V_1 \\
g(V_1) & \sim g(2 \mu_{V_1} - V_1) \\
M_1 V_1 + M_2 & \sim 2 M_1 \mu_{V_1} - M_1 V_1 + M_2 \\
V_2 & \sim  2 (M_1 \mu_{V_1} + M_2)  - (M_1 V_1 + M_2) \\
V_2 & \sim 2 (\mu_{V_2}) - V_2.
\end{align*}

Thus, $V_2$ has a point of symmetry. 
\end{IEEEproof}

\begin{theorem}
 \label{thm::MaintiningSym}
 If $X_{\text{init}}$ and $X_{\text{target}}$ are independent of $w_1^T$ and $v_1^T$, and both have Point Symmetries, then the vectors $X_t$ as well as $X$ will all have Point Symmetries for any matrix $M$.
\end{theorem}

\begin{Pf}
{
 First, note that $w_1^T$ and $v_1^T$ are Gaussian random vectors, and therefore have Point Symmetries across their mean points. Since $X_{\text{init}}$ and $X_{\text{target}}$ are independent of $w_1^T$ and $v_1^T$, then the vector $Q$ also has a Point Symmetry across the mean point (which is the concatenation of the mean points of the respective components of $Q$); we denote this point by $\mu_Q$. Then, by virtue of Lemma~\ref{lem:symmetry_map}, $X$ (respectively $X_t$) will also have Point Symmetry across the point $M \mu_Q$ (respectively across the point $\mu_Q$ left-multiplied by the corresponding section of the matrix $M$).}
\end{Pf}

\textbf{{Revisiting Example-3 of Section~\ref{sec:expected}}}:
{Example 3 in Section~\ref{sec:expected} shows an example where the initial and target points exhibit Point Symmetry. In such an example, the LQR controller is a linear function of the previous states (one can find such a controller by applying the KKT conditions). Since the system is noiseless, then the estimated states are equal to the observations. Therefore, the conditions for Theorem~\ref{thm::MaintiningSym} are met, and Point Symmetry is preserved for $X_t$ and the whole trajectory $X$. Note, however, that the point of symmetry for $X_t$ changes with $t$, {\it i.e.}, it progresses along the $x$ axis as shown in Figure~\ref{fig::simulationMirroring}. }

\section{Optimizing The Worst Case Distortion $D_W$}
\label{sec:worst}


{The expected distortion metric might not be well-suited for some applications (for example if an adversary wants to shoot a drone). In this case, the adversary's estimate needs to be far from the actual state \textit{at all} time instances. Therefore, a more appropriate metric would be to consider the worst case distortion for the adversary. Consider for example the scheme in Fig.~\ref{fig::mirroring}. Here, the adversary's estimate is always the center point and therefore the maximum expected distortion is achieved. However, when the drone is close to the center, its mirror image will also be close to the center.  At this particular time instance, the adversary's distortion will be very small and thus the adversary will essentially know the position. }

In this section, we present an encryption scheme that attempts to maximize 
the worst  case distortion for Eve. 
The main idea is to obfuscate the initial state in such a way that Eve, even if she optimally uses her knowledge about the dynamics and her observations,  her best estimate is close to the maximal distortion. We start by studying the problem of distorting the transmission of a single random variable in Theorems~\ref{thm:gaussian_scalar} and~\ref{thm:gaussian_vector}. These results then form the basis for maximizing the worst case distortion of a trajectory, as described in Theorem~\ref{thm:traj}.


\subsection{Building Step: Scalar Case}
Consider the case where the system wants to communicate a single scalar random variable $X$ to Bob by transmitting $Z$. The worst case distortion  $D_W$ for Eve will be $D_W = \min_{Z} \text{Var}(X|Z)$. Note that if Eve does not overhear $Z$, Eve {uses} the minimum mean square {error} estimate ({\textit{i.e.},} the mean value) as her estimate,
and thus experience a worst case distortion equal to the variance of $X$. 

We first assume that $X \sim \Set{N}(0,1)$, and thus, the worst case distortion can not be larger than $1$ by~\eqref{eq::UpperBoundDW}. We next
develop our scheme progressively, from simple to more sophisticated steps.
We will also use the following lemma.
\begin{lemma}
	\label{lem:var}
{The variance of a Bernoulli random variable taking values $a$ and $b$ with probabilities $p_a$ and $p_b$, respectively, is given by $p_a p_b(a-b)^2$.}
%
\end{lemma}

\textbf{Mirroring.} 
Reflecting around the origin (as we did for optimizing the average case distortion in Section~\ref{sec:expected}) does not work well when $X$ takes small values: indeed $\text{Var}(X|Z)$ is $Pr(X\!=\!Z|Z) (Pr(X\!=\!-Z|Z)) (Z-(-Z))^2$ using Lemma~\ref{lem:var} and has a worst case value that goes to zero as $Z$ approaches zero.
%
%
%

\textbf{Shifting.}
To avoid this, we could try to use a ``shifting" scheme where we add {a constant $\theta$} to $X$ whenever the shared key bit is one; but now this scheme does not perform well for large values of {{$Z$:} as $Z$ increases  $\text{Var}(X|Z)$  goes to zero. This is because using Lemma~\ref{lem:var}:
\begin{align*}
\text{Var}(X|Z) & = Pr(X\!=\!Z|Z) (Pr(X\!=\!Z\!-\!\theta|Z)) (Z\!-\!(Z\!-\!\theta))^2 \\
&= Pr(X= Z|Z) (Pr(X = Z - \theta|Z)) (\theta)^2,
\end{align*}
and $Pr(X= Z|Z) (Pr(X = Z-\theta|Z))$ goes to zero for large value of $Z$. }
%

%

\textbf{Shifting+Mirroring.} 
We here combine shifting and mirroring, in order to achieve a good performance for both small and large values of  $X$. We  start from the case where we have $k=1$ bit of key and then go to the case $k\geq 1$. \\
$\bullet \quad {k=1}.$ We select a $\theta_1\in \mathbb{R}$ that determines a window size ($\theta_1 $ is public and known by Eve). The encoding function is
\begin{align*}
Z & = \mathcal{E} (X, K) = \left\{\begin{array}{cl} X & \text{if } K = 0 \\ -X & \text{if } K = 1, \ |X| > \theta_1 \\ X + \theta_1 & \text{if } K = 1, \ -\theta_1 \leq X < 0 \\ X - \theta_1 & \text{if } K = 1, \ 0 \leq X < \theta_1\end{array} \right.
\end{align*} 
We note that there is one particular value of $X$,  $X = \theta_1$, which we do not transmit. Since this is of zero probability measure, it can be safely ignored.
Given $Z$, there are two possibilities for $X$: 

\begin{align*}
X \in \left\{\begin{array}{cl}
\{Z, -Z\} & \text{if } |Z| > \theta_1 \\
\{Z, Z+\theta_1\} & \text{if } -\theta_1 \leq Z < 0 \\
\{Z, Z-\theta_1\} & \text{if } 0 \leq Z < \theta_1 .
\end{array}\right.
\end{align*}
Using the fact that $X \sim \Set{N}(0,1)$, we can calculate the posterior probabilities $Pr(X|Z)$ and use Lemma~\ref{lem:var}  to compute $\text{Var}{(X|Z)}$. Fig.~\ref{fig:plot_distortion_mirror_shift} plots $\text{Var}(X|Z)$ for $\theta=1.76$. The worst case distortion in this case becomes $0.4477$, which is the best we can hope for if we have only one bit of shared key. {This follows because for any mapping from $X$ to $Z$, a transmitted symbol $Z$ can have at most two pre-images (as Bob needs to reliably decode with one bit of key), and if one of these is $X=0$, then no matter what the second one is, the distortion corresponding to  $Z$ will be at most $0.4477$. Equality occurs when the second pre-image of  $Z$ is $\pm 1.76$. Note that our scheme also maps {$0$} to {$-1.76$} (for $\theta_k = 1.76$).}  \\

\begin{figure}[ht]
	\centering
	\begin{subfigure}[t]{0.45\linewidth}
		\begin{tikzpicture}
		\begin{scope}[scale=0.5]
		\draw (-3.5,0) -- (3.5,0);
		\draw (0,-0.1) node [below] {$0$};
		\draw (2,-0.1) node [below] {$\theta_2$};
		\draw (-2,-0.1) node [below] {$-\theta_2$};
		\draw (-2,-0.1) -- (-2,0.5);
		\draw (2,-0.1) -- (2,0.5);
		\draw (0,-0.1) -- (0,0.5);
		\draw (0,-0.1) -- (0,0.5);
		\draw (1,-0.1) -- (1,0.5);
		\draw (-1,-0.1) -- (-1,0.5);
		\draw (0.6,-0.2) rectangle (1, 0.2);
		\draw[fill=black] (-0.4,-0.2) rectangle (0, 0.2);
		\draw (3,0) circle(3mm);
		\draw[fill=black] (-3,0) circle(3mm);
		\end{scope}
		\begin{scope}[scale=0.5, shift={(0,-4)}]
		\draw (-3.5,0) -- (3.5,0);
		\draw (0,-0.1) node [below] {$0$};
		\draw (2,-0.1) node [below] {$\theta_2$};
		\draw (-2,-0.1) node [below] {$-\theta_2$};
		\draw (-2,-0.1) -- (-2,0.5);
		\draw (2,-0.1) -- (2,0.5);
		\draw (0,-0.1) -- (0,0.5);
		\draw (0,-0.1) -- (0,0.5);
		\draw (1,-0.1) -- (1,0.5);
		\draw (-1,-0.1) -- (-1,0.5);
		\draw (0.6,-0.2) rectangle (1, 0.2);
		\draw[fill=black] (-1.4,-0.2) rectangle (-1, 0.2);
		\draw (3,0.1) circle(3mm);
		\draw[fill=black] (3,-0.1) circle(3mm);
		\end{scope}
		\end{tikzpicture}		
		\caption{}
		\label{fig:scheme}
	\end{subfigure}
	\begin{subfigure}[t]{0.45\linewidth}
		\centering
		\includegraphics[width=1.5in]{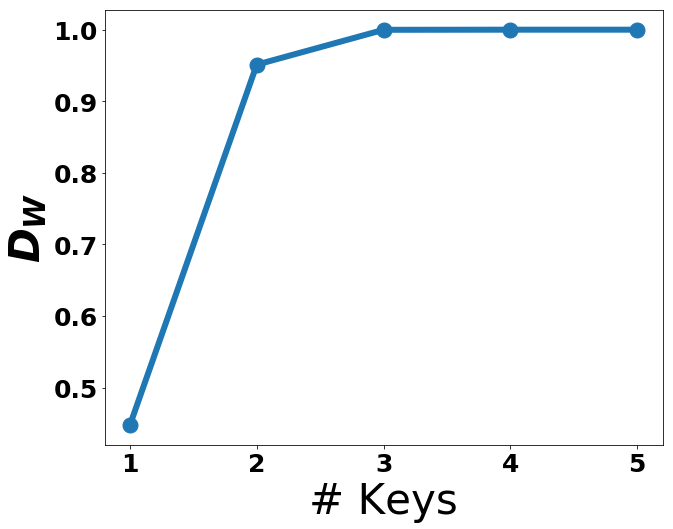}
		\caption{}
		\label{fig:dist_vs_keys}
	\end{subfigure}
	\caption{(a) Transparent shapes represent true values and solid shapes represent their respective mapping when two bit key is $11$ and $10$ respectively. (b) $D_W$, as a function of number of keys for optimal choice of $\theta_k$.}		
\end{figure}
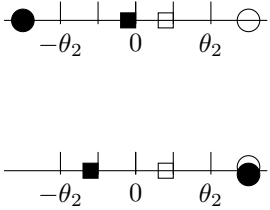
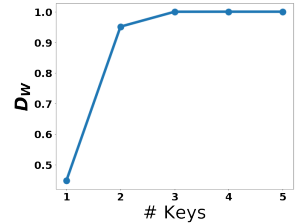	
\begin{figure}[ht]
	\centering
	\includegraphics[width=3.0in]{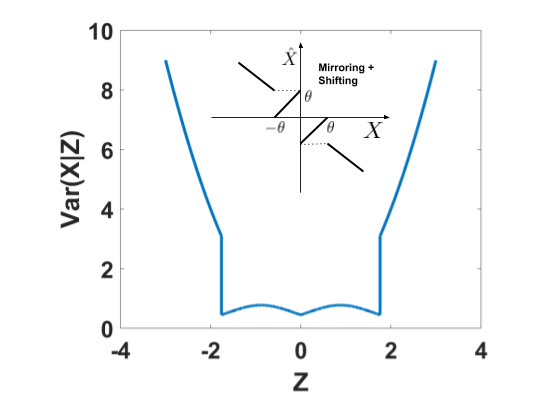}
	\caption{$\text{Var}(X|Z)$ versus Z for the shifting+mirroring scheme with $\theta_1=1.76$; $D_W = 0.4477$. $\text{Var}(X|Z)$ is $Z^2$ for $|Z| > \theta_1$ and is $ p Z^2 + (1-p) \tilde{Z}^2  - (p Z + (1-p) \tilde{Z})^2$ for $|Z| \leq \theta_1$, where $p = f(Z)/\left(f(Z) + f(\tilde{Z})\right)$, with $f(Z) \sim \mathcal{N} (0,1)$, and $\tilde{Z} =  Z + \theta_1  \text{mod } [-\theta_1, \theta_1)$. }
	\label{fig:plot_distortion_mirror_shift}
\end{figure}
\noindent $\bullet \quad k\geq 1$. For $K \in \{0,1\}^k$,
we use the following encoding:
\begin{align}
Z & = \mathcal{E} (X, K) \\
& = \left\{\begin{array}{cl}
\left\{\begin{array}{cl} X & \text{if } K < 2^{k-1}  \\ -X & \text{if } K \geq 2^{k-1} \end{array}
\right. & |X| > \theta_k \\
X + K \frac{2\theta_k}{2^{k}} \text{mod } [-\theta_k, \theta_k) &  X \in [-\theta_k, \theta_k),\nonumber
\end{array}
\right.
\end{align} 
where the optimal value of  the constant $\theta_k$ depends on the number $k$ of keys we have, $K$ is the decimal equivalent of a binary string of length $k$, and $r \text{mod } [a,b) = r - i(b-a) $ is such that $i$ is an integer and $r - i(b-a) \in [a,b)$ for $r,a,b \in \mathbb{R}$. Intuitively, if $|X|> \theta_k$ then for half of the keys, we reflect across origin and for other half we do nothing; if $|X| < \theta_k$, we divide this window of size $2\theta_k$ into ${2^k}$ equal size windows and shift a point from one window to another by jumping $K$ (in decimal) windows. An example for $k=2$  is shown in Fig.~\ref{fig:scheme} for the key values $K=11$ and $K=10$. Fig.~\ref{fig:dist_vs_keys}  plots $D_W$ as a function of the number of keys $k$. Using $k=3$ and $\theta_3=4.84$ we achieve $D_W=0.9998$ which is very close to $1$, the best we can hope for.

\textbf{Remark:} {We optimize the parameter $\theta_k$ of our scheme assuming Gaussian distribution. In particular, $\theta_k  = \arg\max_{\theta_k}  \left( \min_Z  D_W(Z) \right)$ and $D_W(Z)$ is $Z^2$ if $|Z| > \theta_k$, and is $ \sum_{K \in \{0,1\}^k} \left( p_K \left(Z_K\right)^2 \right) - \left(\sum_{K \in \{0,1\}^k} p_K Z_K \right)^2$, if $|Z| \leq \theta_k$.
	Here, $p_K={f (Z_K)}/{\sum_{L \in \{0,1\}^k} f(Z_L)   }$ with  $f(Z_K) \sim \mathcal{N} (0, 1)$, and $Z_K$ for $K \in \{0,1\}^k$ is defined as $Z_K  = Z + K_d \frac{2\theta_k}{2^k} \ \text{mod} \ [-\theta_k, \theta_k]$,
	with $K_d$ being the decimal equivalent of $K$.	
	
	We pick a choice of $\theta_k$ by computationally iterating over the values of $\theta_k$ to find one which maximizes $D_W$, \textit{i.e.}, $\theta_k = \arg \max_{\theta_k} \min_{Z} (D_W (Z))$.  	
}

For other distributions, the optimal choice of $\theta_k$ and the corresponding worst case distortion would be different.

\begin{theorem}
	\label{thm:gaussian_scalar}
	A Gaussian random variable with mean $\mu$ and variance $\sigma^2$ can be near perfectly ($\sim0.9998$ times the perfect distortion) distorted in worst case settings by just using three bits of shared keys.
\end{theorem}
\begin{IEEEproof}
Generate the random variable  $V \sim \Set{N}(0,1)$ as $V = (X - \mu)/\sigma$ and encrypt it using $k=3$ key bits and the previously described scheme. 
{We transmit the mean $\mu$ and the variance $\sigma^2$ uncoded, and show that near perfectly distorting the standard Gaussian $V$ results in near perfect distortion of $X$ for Eve.}
For  $c = 0.9998$ we have 
\begin{align*}
{D_W} & = \min\limits_Z \text{Var}(X|Z)  = \min\limits_Z \text{Var}(\sigma V + \mu |Z) \\
& = \sigma^2 \min\limits_Z \text{Var}(V|Z)  = c\sigma^2.
\end{align*} 
\end{IEEEproof}

\subsection{Vector Case and Time Series}
\begin{theorem}[Proof in Appendix~\ref{app:gaussian_vector}]
	\label{thm:gaussian_vector}
	For a Gaussian random vector $X \in \mathbb{R}^n$ with mean $\mu$ and a diagonal covariance matrix $\Sigma$ we can achieve $D_W$ within $0.9998$ of the optimal by using $3n$ bits of shared keys.
\end{theorem} 
This theorem uses our 3-bit encryption for each element {in} the vector. Assume now that this vector captures the probability distribution of the initial state of dynamical system; by encrypting this state we can guarantee the following. 
{ 
\begin{theorem}[Complete Proof in Appendix~\ref{app:traj}]
  \label{thm:traj}
  Using $3n$ bits of shared keys, the shifting+mirroring scheme achieves $D_W \geq c \cdot \text{tr}\left( \left| \Lambda \right|^{2t} \Sigma \right)$ with $c = 0.9998$ for the dynamical systems~\eqref{eq:sys_model} with $C = I$, $v_t = 0$, the singular value decomposition of $A$ is $A = \Phi \Lambda V^H$, and initial state $X_1 \sim \mathcal{N}(\mu, \Sigma)$, where $\Sigma$ is diagonal covariance matrix, and $U_t$ and $w_t$ are independent of $X_t$. Moreover, if $|\lambda_i| \geq 1, \forall i$, where $\lambda_i$ is the $i$-th singular value of $A$, then $D_W$ is within $0.9998$ of the maximum distortion.
\end{theorem}
}
{\textbf{Remark}: {Although the independence assumption on the inputs is rather restrictive, the result}
serves as a stepping stone {towards understanding general cases}}.

\begin{IEEEproof}
 The system transmits $ Z_1  = f(Y_1, K) = f(X_1,K)$ where $f$ is the encoding in Theorem~\ref{thm:gaussian_vector}, and {for $t \in [T-1]$,
	\begin{align*}
	Z_{t+1} = A Z_t + (Y_{t+1} - A Y_{t}) 	
	= A Z_t + B U_t + w_t. 
	\end{align*}
	}
 Bob can  decode $X_1$ using  $Z_1$ and $K$.  Then:
	\begin{align*}
	\hat{X}_{t+1} & = Z_{t+1} - A Z_t + A \hat{X_t} \\
	& = (A Z_t + B U_t + w_t) - A Z_t + A \hat{X_t} \\
	& = A X_t  + BU_t + w_t = X_{t+1} , \ \ \ \forall t \in[T-1].
	\end{align*}
	Eve's distortion is calculated  in the Appendix~\ref{app:traj}. 
	 \end{IEEEproof}

\noindent\textbf{Complexity:} $\mathcal{O}(n^2)$ per time instance for both encoding and decoding.

{
\noindent\textbf{Case study}: We take three choices of $A$ of sizes $2 \times 2$, first having all singular values no smaller than one, in particular $[1.01, 1]$, second having singular values $[1.5, 0.5]$ and third having singular values of $[0.8, 0.9]$. For a given co-variance matrix $\Sigma = [2, 0; 0, 3]$ for the initial state, we plot the evolution of distortion at the adversary's end corresponding to our encryption scheme and compare with the $c \text{tr}(\Sigma)$. This evolution is shown in Fig.~\ref{fig:case_study}. As we can observe, when $A$ has all the singular values more than one, the distortion at adversary's end is always at least $c \text{tr} (\Sigma)$, whereas for other cases in eventually goes to zero.
}

\begin{figure}[ht]
	\centering
	\includegraphics[width=0.75\linewidth]{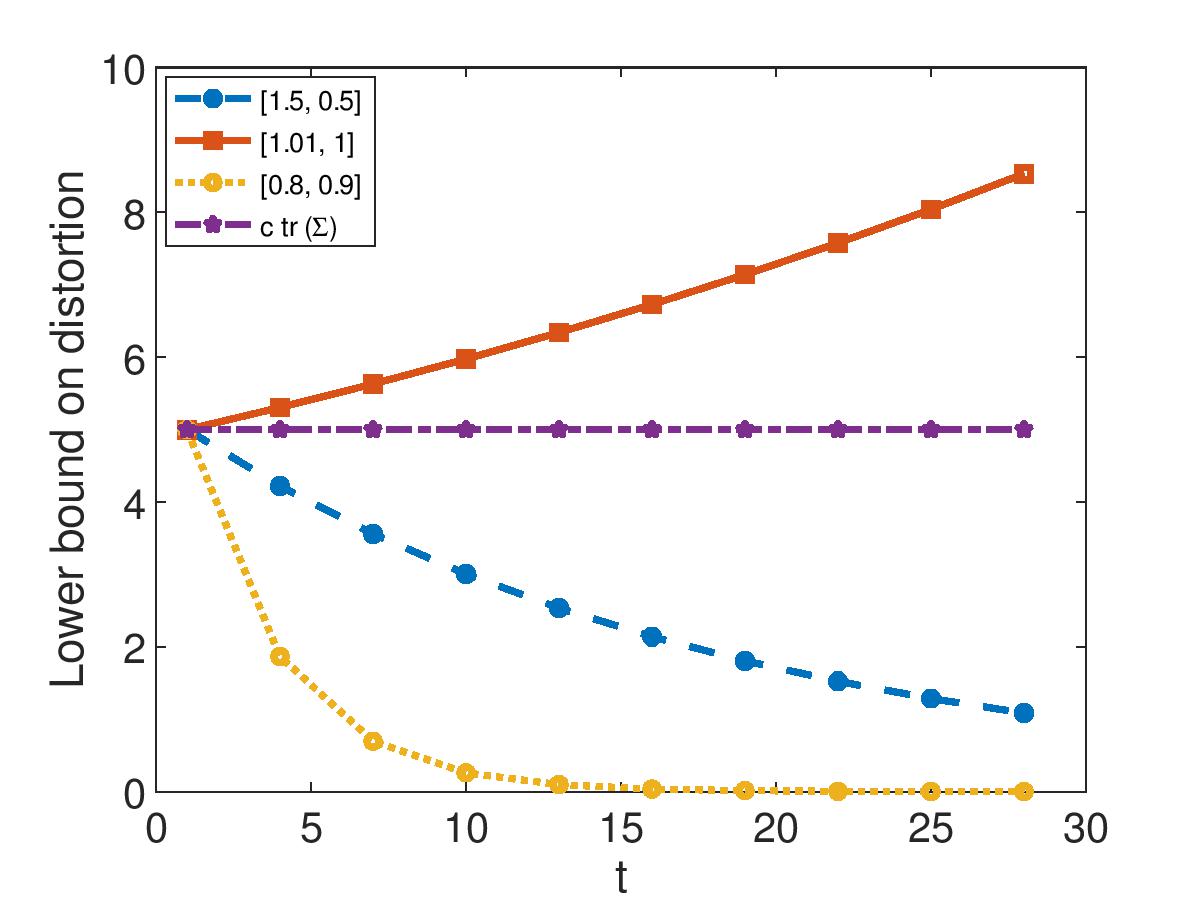}
	\caption{Evolution of the distortion for Eve for shifting+mirroring based scheme}
	\label{fig:case_study}
\end{figure}

\section{{Distorting the Inputs and the Implication on States}}
\label{sec:inputs}

In many situations, it is easier to obtain a handle on the distribution of the input sequence than on the distribution of the state transition sequence.
{{Moreover, in some situations, a simple transformation of the original trajectory would lead to a fake trajectory that does not obey the system dynamics and thus can be detected by the adversary -- alternatively, the state trajectory distribution does not have useful symmetry properties. }
Motivated by these, here we consider a different setup where Alice encodes and transmits the input sequence to Bob instead of the state transition sequence, {{{\it i.e.}, $Z_t \coloneqq \mathcal{E}_t (U_t, K)$}. Under this setup, using the mirroring based scheme on inputs, one can provide guarantees on the level of {{average/worst-case} distortion for Eve's estimate of the inputs. {{We} then ask the following question: if Alice encodes and transmits the input vectors, how does the guarantees on average and worst-case distortions on the \textbf{inputs} translates to the guarantees on average and worst-case distortions on the \textbf{states}?

%

{Formally, we} consider the system model in~\eqref{eq:sys_model} with zero noise, {{\it i.e.}, $w_t = 0$. Following the definition of average and worst-case distortions in~\eqref{eq:average_case} and~\eqref{eq:worst_case} respectively, the {distortions} on inputs and states vectors {are} as follows:\\
	
\noindent Expected distortions:
{\begin{align*}
D_E^{(X)} &= \frac{1}{T} \mathbb{E}_Z \sum_{t=1}^{T} \text{tr} \left( R_{X_t|Z} \right), & D_E^{(U)} = \frac{1}{T} \mathbb{E}_Z \sum_{t=0}^{T-1} \text{tr} \left( R_{U_t|Z} \right).
\end{align*}
}
Worst-case distortions:
{\begin{align*}
D_W^{(X)} &= \min_Z \min_{t} \text{tr} \left( R_{X_t|Z} \right), &D_W^{(U)} = \min_Z \min_{t} \text{tr} \left( R_{U_t|Z} \right).
\end{align*}
}
{The following Theorem provides a relation between the distortion on the input vector and the distortion on the state vector for both expected and worst-case scenarios. In particular, it provides relations between {$D_E^X$, $D_E^U$, and $D_W^X$, $D_W^U$} .}
{
\begin{theorem}
\label{thm::MirroringPerformanceCausal_input}
If the input vectors $U_t, \: t \in [T-1] \cup \{0\}$ 
satisfy the following condition, 
{
\begin{equation}
\label{eq::MirroringPerformanceCausal_input_cond}
 \mathbb{E}_Z \left[ \text{tr}\left( \underbrace{\sum\limits_{i=1}^t \sum\limits_{j=i+1}^t  B^\prime (A^{t-j})^\prime A^{t-i} B R_{U_{i-1}U_{j-1}|Z} }_{\Phi_t} \right) \right]\geq 0,
\end{equation}
}%
then, for given distortions $D_E^{(U)}$ and $D_W^{(U)}$, the following bounds holds:
{
\begin{align*}
D_E^{(X)} &\geq \lambda_{\min} \left(B^\prime B \right) D_E^{(U)},
&D_W^{(X)} \geq \lambda_{\min} \left(B^\prime B \right) D_W^{(U)},
\end{align*}
}
where $\lambda_{\min} \left(B^\prime B \right)$ is the minimum eigenvalue of $B^\prime B$.
\end{theorem}

Theorem~\ref{thm::MirroringPerformanceCausal_input} gives a lower bound on the distortion level of the state vectors when distorting the inputs. The bound holds when the condition~\eqref{eq::MirroringPerformanceCausal_input_cond} holds. {{Examples} where condition~\eqref{eq::MirroringPerformanceCausal_input_cond} holds {are} open loop control systems where the distribution on {{the inputs has} a point of symmetry. We expand more on {{condition}~\eqref{eq::MirroringPerformanceCausal_input_cond} after {{the proof of Theorem~\ref{thm::MirroringPerformanceCausal_input}}.

\begin{IEEEproof}
We start by introducing the following notation $Z = Z_0^{T-1}$, $U = U_0^{T-1}$, $X = X_1^T$. Moreover, without loss of generality, we assume that $X_0  = 0$. The states of the noiseless dynamical system can be expressed as {$X_t = \sum\limits_{i=1}^t A^{t-i}BU_{i-1}$.}
Then we can write
{
\begin{align}
R_{X_t|Z} & = \sum_{i=1}^{t} A^{t-i} B R_{U_{i-1}|Z} B^\prime (A^{t-i})^\prime + 2\Phi_t \label{eq:R_x_t_U_to_X}
\end{align}}

%
{Therefore, combining \eqref{eq::MirroringPerformanceCausal_input_cond} and~\eqref{eq:R_x_t_U_to_X}}, we express $D_E^{(X)}$ as
\begin{align*}
&D_E^{(X)} \geq \dfrac{1}{T}  \mathbb{E}_Z \sum\limits_{t=1}^{T}  \sum\limits_{i=1}^t \text{tr}\left( A^{t-i} B  R_{U_{i-1}|Z}  B^\prime (A^{t-i})^\prime \right) \\
&\stackrel{(a)}{\geq} \dfrac{1}{T}  \mathbb{E}_Z \sum\limits_{t=1}^{T} \text{tr}\left( B  R_{U_{t-1}|Z}  B^\prime \right) = \dfrac{1}{T}  \mathbb{E}_Z \sum\limits_{t=1}^{T}  \text{tr}\left(B^\prime B  R_{U_{t-1}|Z} \right), \\
&\geq \lambda_{\min}(B^\prime B) \dfrac{1}{T}  \mathbb{E}_Z \sum\limits_{t=1}^{T} \text{tr}\left( R_{U_{t-1}|Z} \right) =\lambda_{\min}(B^\prime B) D_E^{(U)},
\end{align*}
\noindent where $(a)$ follows by noting that the matrices $B^\prime \left(A^{t-i} \right)^\prime R_{U_{i-1}|Z} A^{t-i} B$ are positive semidefinite, and therefore their trace is greater than or equal to zero. To see that they are indeed positive semidefinite, note that $R_{U_{i-1}|Z}$ is positive semidefinite, therefore it has the eigen-decomposition $R_{U_{i-1}|Z} = \Sigma \Lambda \Sigma^\prime$. Therefore, the claim follows by noting that, for any vector $z$, we have
\begin{align*}
&z ^\prime B^\prime \left(A^{t-i} \right)^\prime R_{U_{i-1}|Z} A^{t-i} B z = \\
& z^\prime B^\prime \left(A^{t-i} \right)^\prime \Sigma \Lambda^{1/2} \Lambda^{1/2} \Sigma^\prime A^{t-i} B z = \| \Lambda^{1/2} \Sigma^\prime A^{t-i} B z \|^2 \geq 0.
\end{align*}
{Identical arguments can be made to show the bound on $D_W^{(X)}$.}
\end{IEEEproof}

{Next, we show some sufficient conditions which ensure that condition~\eqref{eq::MirroringPerformanceCausal_input_cond} holds:}\\

\noindent {\it 1) $U_{i}$ and $U_{j}$ are uncorrelated for $i \neq j$ and $U := U_0^{T-1}$ has Point Symmetry:}
In this case, the optimal mirroring scheme is to mirror the point $U$ across the point of symmetry. Therefore, given $Z$, $U_i$ takes two values: $Z_i$ with probability $p_Z$ and $\tilde{Z_i}$ with probability $1-p_Z$, where $p_Z$ is equal to $p_Z  = {f_U(Z)}/(f_U(Z) + f_U(\tilde{Z})) = 0.5$,
which follows from {the Point Symmetry assumption on $U$. {So we have $\mathbb{E} [U_i | Z] = \frac{Z_i +\tilde{Z}_{i}}{2}$.} Therefore,} $R_{U_{i}U_{j}|Z}$ can be computed as follows 
{
\begin{align*}
R&_{U_{i} U_{j} | Z } = \mathbb{E}_{U|Z} \left[ \left(U_i - \mathbb{E} [U_i |Z] \right)  \left(U_j - \mathbb{E} [U_j |Z] \right) ^\prime \right]\\
& =  \frac{1}{4} \left(Z_i - \tilde{Z}_i \right) \left(Z_j - \tilde{Z}_j \right)^\prime = \left(Z_i - \mu_{Z_i} \right) \left(Z_j - \mu_{Z_j} \right)^\prime.
\end{align*}
}

{Then we have $E_Z R_{U_i U_j | Z }= R_{U_i U_j} = 0$ by noting that $Z_i, Z_j$ have {the} same {distribution as $U_i, U_j$:}
\begin{align*}
&f_{Z_i, Z_j}(z_i, z_j)  = \frac{1}{2} \left( f_{U_i, U_j} (z_i, z_j) +  f_{U_i, U_j} (\tilde{z}_i, \tilde{z}_j) \right), \\
&= \frac{1}{2} \left( f_{U_i} (z_i) f_{U_j} (z_j) +  f_{U_i} (\tilde{z}_i) f_{U_j} (\tilde{z}_j) \right) =  f_{U_i, U_j} (z_i, z_j).
\end{align*} }

\noindent {\it 2) $A$ and $R_{U_i, U_j | Z}$ are positive semidefinite matrices for all $i$, $j$ and $Z$:} this follows because, if $A$ is positive semidefinite, then so is $A^i$ for any value of $i$. Therefore we can write

{
\begin{align*}
& \text{tr}\left( B^\prime (A^{t-j})^\prime A^{t-i} B R_{U_{i-1}U_{j-1}|Z} \right) \\ 
&= \text{tr}\left( B^\prime A^{t-j} A^{t-i} B R_{U_{i-1}U_{j-1}|Z} \right) \\
&= \text{tr}\left( B^\prime A^{2t-i-j} B R_{U_{i-1}U_{j-1}|Z} \right)\\
&= \text{tr}\left( B^\prime \Sigma \Lambda^{1/2} \Lambda^{1/2} \Sigma^\prime B R_{U_{i-1}U_{j-1}} \right) \\
&= \text{tr}\left( \Lambda^{1/2} \Sigma^\prime B R_{U_{i-1}U_{j-1}|Z} B^\prime \Sigma \Lambda^{1/2} \right) \\
&= \text{tr}\left( \Lambda^{1/2} \Sigma^\prime B \Phi \Gamma^{1/2} \underbrace{\Gamma^{1/2} \Phi^\prime B^\prime \Sigma \Lambda^{1/2}}_{D} \right) = \text{tr}\left( D^\prime D \right) \geq 0.
\end{align*}}

{
\section{Conclusion}
In this work, we considered distortion-based security for CPSs as a complementary security approach which optimizes an alternative security goal. This approach for security is suitable for CPS applications where the estimation of the adversary about the states is required to be "far" from the actual state value. We provided security schemes which aim to optimize for both the average and worst-case distortion. For the average distortion, we showed the surprising result that $1$-bit schemes are optimal for certain distributions. We then provided the expression for the attained level of distortion for a general security scheme. For worst-case distortion, we considered an initial situation where we proposed an encryption scheme which achieves near optimal distortion.
}

\section{Appendices}

\subsection{Proof of Theorem~\ref{thm::MirroringPerformanceCausal} and Corollary~\ref{cor::SymmetricDist} }
\label{app::MirroringPerformanceCausal}
%

We start by computing $R_{X_t|Z_1^T}$. Note that given a sequence of transmitted symbol $Z_1^T$ there are two possible values of sequence of message symbols $X_1^T$ which are $X_1^T = Z_1^T$ and $X_1^T = \tilde{Z}_1^T$, where $\tilde{Z_t}$ is $\alpha_t^{-}(Z_t)$ {and $\tilde{X_t}$ is $\alpha_t^{-}(X_t)$}.

{The posterior probability of $X_t=Z_t$ given $Z_1^T$ \textit{i.e.}, $Pr (X_t = Z_t |Z_1^T)$ will be equal to {$Pr (X_1^T = Z_1^T|Z_1^T) \coloneqq p_Z$. We note that $p_Z = \frac{f(Z)}{f(Z) + f(\tilde{Z})}$}, where $\tilde{Z}: = [{\tilde{Z}_1}^\prime \:\: {\tilde{Z}_2}^\prime \:\: \cdots \:\: {\tilde{Z}_T}^\prime]^\prime$.} Then, 
{$\mathbb{E}(X_t|Z_1^T) = p_Z {Z_t} + (1-p_Z)(\tilde{{Z_t}})$.}
With this,
\begin{align*}
R_{X_t|Z_1^T} &= \mathbb{E}_{X_t|Z_1^T} \left[ \left(X_t - \mathbb{E}(X_t|Z_1^T) \right)  \left(X_t - \mathbb{E}(X_t|Z_1^T) \right)^\prime \right]\\
&= p_Z (1-p_Z)^2 (Z_t - \tilde{Z_t})(Z_t - \tilde{Z_t})^\prime \\
& \quad + (1-p_Z) p_Z^2 (Z_t - \tilde{Z_t})(Z_t - \tilde{Z_t})^\prime \\
& = p_Z (1-p_Z)(Z_t - \tilde{Z_t})(Z_t -\tilde{Z_t})^\prime \\
D_E &= \frac{1}{T} \mathbb{E}_Z \sum\limits_{t=1}^T \text{tr}\left(R_{X_t|Z_1^T} \right) \\
&= \frac{1}{T} \mathbb{E}_Z \sum\limits_{t=1}^T \text{tr}\left(p_Z (1-p_Z)(Z_t - \tilde{Z_t})(Z_t - \tilde{Z_t})^\prime\right) \\
&= \frac{1}{T} \mathbb{E}_Z \sum\limits_{t=1}^T p_Z (1-p_Z) \text{tr}\left( (Z_t - \tilde{Z_t})(Z_t - \tilde{Z_t})^\prime\right) \\ 
\end{align*}
\begin{align*}
&= {\frac{1}{T} \mathbb{E}_Z  \sum\limits_{t=1}^T p_Z (1-p_Z)  \|Z_t - \tilde{Z_t} \|^2} \\
&= {\frac{1}{T} \mathbb{E}_Z \sum\limits_{t=1}^T  \frac{f_X(Z) f_X(\tilde{Z})}{(f_X(Z) + f_X(\tilde{Z}))^2}  \|Z_t -\tilde{Z_t} \|^2.}
\end{align*}

Now, $Z_1^T$ is the transmitted symbols if $X_1^T=Z_1^T$ and key was zero or if $\{X_t=\tilde{Z_t}, \ \forall t \in [T] \}$ and key was one. So $f_Z(Z) = \frac{f_X(Z) + f_X(\tilde{Z})}{2}$. Thus $D_E$,
\begin{align*}
&= \frac{1}{T} \mathbb{E}_Z \sum\limits_{t=1}^T  \frac{f_X(Z) f_X(\tilde{Z})}{(f_X(Z) + f_X(\tilde{Z}))^2}  \|Z_t - \tilde{Z_t} \|^2 \\
&= \frac{1}{T}\int\!f_Z(Z) \sum\limits_{t=1}^T \frac{f_X(Z) f_X(\tilde{Z})}{(f_X(Z) + f_X(\tilde{Z}))^2}  \|Z_t - \tilde{Z_t} \|^2 \!dZ \\
&= \frac{1}{2T}\int \sum\limits_{t=1}^T \frac{f_X(Z) f_X(\tilde{Z})}{f_X(Z) + f_X(\tilde{Z})} \|Z_t - \tilde{Z_t}\|^2 \ dZ \\&= \frac{1}{2T} \mathbb{E}_{X} \sum\limits_{t=1}^T \frac{f_X(\tilde{X})}{f_X(X) + f_X(\tilde{X})} \|Z_t- \tilde{Z_t}\|^2 
\end{align*}

\begin{align*}
&= \frac{1}{2T} \mathbb{E}_{X} \sum\limits_{t=1}^T \frac{f_X({\alpha^{-}}(X))}{f_X(X) + f_X({\alpha^{-}} (X))} \|X_t- {\alpha_t^{-}}(X_t)\|^2,
\end{align*}


which proves~\eqref{eq::onebit_dist}. 
Again, if we can choose $S_t$'s, $b_t$'s where $\alpha_t()$ is mirroring across planes given by $S_t x = b_t$ such that,
\begin{align*}
f_X(X) = f_X(\alpha^{-1}(X)), \ \forall X \in \mathbb{R}^{n T},
\end{align*} 
the distortion $D_E$ becomes,
\begin{align*}
D_E & = \frac{1}{4T} \mathbb{E}_X \sum\limits_{t=1}^T \|X_t - {\alpha_t^{-}}(X_t) \|^2
{\stackrel{(a)}{=}} 
\frac{1}{T} \sum\limits_{t=1}^T \mathbb{E}_{X_t} \|S_tX_t - b_t \|^2 \\
& =  \frac{1}{T} \sum\limits_{t=1}^T \text{tr} \left( S_t R_{X_t} S_t^\prime + (b_t - S_t \mu_{X_t}) (b_t - S_t \mu_{X_t})^\prime\right),
\end{align*}
where (a) follows as $\alpha_t(.)$ is mirroring across plane given by $S_tx = b_t$, and thus ${\alpha_t(x) = \alpha_t^{-1}} (x) = (I - 2 S_t^\prime S_t) X_t + 2 S_t^\prime b_t $. This proves \eqref{eq::DEE_simple_causal}. 
{\subsection{Proof of Theorem~\ref{thm::MirroringPerformanceCausal_multibit}}
\label{app::multipleKproof}
Since given $Z$, there are $2^k$ possibilities of $X_1^T$; $X_1^T = \alpha^{-1 (K)} (Z),  K \in [0:2^k-1] $,
we start by computing, 
\begin{align*}
 p^{(K)}_Z &\coloneqq Pr(X_t = \alpha_t^{-(K)}(Z_t) |Z ) = Pr(X = \alpha^{-(K)}(Z) |Z ) \\
 &= \frac{1}{f_Z(Z)}Pr(Z|X = \alpha^{-(K)}(Z) ) f_X(\alpha^{-(K)}(Z)) \\
 &\stackrel{(a)}{=} \frac{f_X(\alpha^{-(K)}(Z))}{\sum\limits_{j=0}^{2^k-1} f_X(\alpha^{-(j)}(Z))}, \ K \in [0:2^k-1], 
\end{align*}
where $(a)$ follows by noting that $Pr(Z|X = \alpha^{-(K)}(Z)|Z)$ is equal to the probability of the key being equal to $K$, which is $1/2^k$. {Let $S =\sum\limits_{j=0}^{2^k-1} f_X(\alpha^{-(j)}(Z))$. Then $\mathbb{E}(X_t|Z)$ equals 
\begin{align*}
\sum\limits_{K=0}^{2^k-1} \alpha^{-(K)}(Z_t) p_Z^{(K)} = \frac{1}{S} \sum\limits_{K=0}^{2^k-1} \alpha^{-(K)}(Z_t) f_X(\alpha^{-(K)}(Z)).
\end{align*}}
We can then compute $\text{tr}\left(R_{X_t|Z} \right)$ as,
\begin{align*}
\mathbb{E}_{X_t|Z} {\left\|X_t \!-\! \mathbb{E}(X_t|Z) \right\|^2} = \frac{1}{S^3} \sum\limits_{K=0}^{2^k-1} f_X\left(\alpha^{-(K)}(Z)\right) \left\| R_{t}^{(K)}  \right\|^2,
\end{align*}
where $R_{t}^{(K)}\!\!=\!\sum\limits_{\ell=0}^{2^k-1} f_X(\alpha^{-(\ell)}(X)) \left( \alpha_t^{-(\ell)}(X_t) - \alpha_t^{-(K)}(X_t) \right)$.
Plugging $tr\left(R_{X_t|Z} \right)$ in the expression of $D_E$ gives ~\eqref{eq::multibit_dist}. Moreover, if condition~\eqref{eq::sym_cond_general_multibit} is met,~\eqref{eq::multibit_dist} simplifies to~\eqref{eq::kbit_dist_sym}.
}

\subsection{Proof for Theorem~\ref{thm:gaussian_vector}}
\label{app:gaussian_vector}
Let the shared key $K$ is $(K_1, K_2, \ldots, K_n)$ where all $K_i$'s are i.i.d. and uniformly distributed in $\{0,1\}^3$.
Let us also assume that $X = (X^{(1)}, X^{(2)}, \ldots, X^{(n)})$, where each $X^{(i)} \in \mathbb{R}$. Similar to the scheme for scalar case, we create a random vector $V = (V^{(1)}, \ldots, V^{(n)})$ where {$V^{(i)} = (X^{(i)} - \mu^{(i)})/\sqrt{\Sigma_{ii}}$,}
and encode $V^{(i)}$ using key $K_i$ as in the case of a scalar for all $i \in [n]$. Thus, the distortion ${D_W} $ will be,
\begin{align*}
D_W & =\min\limits_{Z} \text{tr}(R_{X | Z})  = \min\limits_{Z} \sum\limits_{i=1}^{n} \text{Var}(X^{(i)} | Z) \\
& = \min\limits_{Z} \sum\limits_{i=1}^{n} (\Sigma_{ii}) \text{Var}(V^{(i)} | Z) = \sum\limits_{i=1}^{n} (\Sigma_{ii}) \min\limits_{Z} \text{Var}(V^{(i)} | Z) \\
& = \sum\limits_{i=1}^{n} (\Sigma_{ii}) \min\limits_{Z^{(i)}} \text{Var}(V^{(i)} | Z^{(i)}) = c \sum\limits_{i=1}^{n} (\Sigma_{ii}) = c~\text{tr} (\Sigma),
\end{align*}
where $c = 0.9998$. And since $\text{tr} (\Sigma)$ is the expected distortion even when the adversary has no observations, and as we can not beat this by~\eqref{eq::UpperBoundDW}, this is optimal.
\subsection{Proof for Theorem~\ref{thm:traj}}
\label{app:traj}
\noindent\textbf{Distortion at the adversary's end.} Based on the coding scheme we can see that the adversary get $BU_t + w_t$ by just subtracting $AZ_t$ from $Z_{t+1}$ for $t \in [1:T-1]$. So the adversary's information is given by following set:
\begin{align*}
E_{\text{info}} & = \left\{Z_1, B U_t + w_t, \ t \in [1:T-1] \right\} \\
&= \left\{f(X_1, K), B U_t + w_t, \ t \in [1:T-1] \right\}.
\end{align*} 

{
Thus, $D(t,Z_1^T) = D(t, E_{\text{info}})  = \text{tr}(R_{{X_t}| E_{\text{info}}}).$ Next, we can write

\begin{align*}
D&(t+1,Z_1^T) = \text{tr}(R_{X_{t+1}|E_{\text{info}}}) = \text{tr}(R_{{(AX_t + B U_t + w_t)}| E_{\text{info}}}) \\
&\stackrel{(a)}{=}\text{tr}(R_{A^{t}X_{1}|E_{\text{info}}}) \stackrel{(b)}{=} \text{tr}(R_{A^{t}X_{1}|f(X_1,K)}) \\
&= \text{tr}(A^t R_{X_1|E_{\text{info}}} (A^t)^\prime) = \text{tr}((A^t)^\prime A^t R_{X_1|E_{\text{info}}}) \\
& \stackrel{(c)}{\geq} c \cdot \text{tr}\left( (A^t)^\prime A^t \Sigma \right) = c \cdot \text{tr}\left( (A^t)^H A^t \Sigma \right) \\
& \stackrel{(d)}{=} c \cdot \text{tr}\left( V (\Lambda^H)^t \Lambda^t V^H \Sigma \right) = c \cdot \text{tr}\left( V \left|\Lambda \right|^{2t} V^H \Sigma \right) \\
&= c \cdot \text{tr}\left( \left|\Lambda \right|^{2t} V^H \Sigma V\right) \stackrel{(e)}{=} c \cdot \text{tr}\left( \left| \Lambda \right|^{2t} \Sigma \right) \stackrel{(f)}{\geq} = c \cdot \text{tr}(\Sigma)
\end{align*}
where $(a)$ follows by noting that $E_{\text{info}}$ contains $BU_t + w_t, \forall t \in [1:T-1]$; 
$(b)$ follows because $U_t$ and $w_t$ are independent on $X_t$;
$(c)$ follows because $\left(A^t\right)^\prime A^t$ is a positive semi-definite matrix and $ R_{{X_1}| E_{\text{info}}}$ being a diagonal matrix with the diagonal entries being element vise greater than $c \Sigma$;
$(d)$ follows by writing the singular value decomposition of $A$ as $A = \Phi \Lambda V^H$;
$(e)$ follows by noting that $\lambda_i$ is the $i$-th singular value of $A$, and $V$ is a unitary matrix;
$(f)$ follows for dynamic systems where $|\lambda_i| \geq 1$.
}

}

\bibliographystyle{IEEEtran}
\bibliography{root} 

\end{document}